\documentclass[dvips]{article}
\usepackage{supertabular,lscape,epsfig}
\usepackage{amssymb}
\usepackage{amsmath}

\DeclareSymbolFont{ppa}{OT1}{ppl}{m}{it}
\DeclareMathSymbol{\vv}{\mathalpha}{ppa}{'166}

\begin{document}

\newcommand{\dd}{\,{\rm d}}
\newcommand{\ie}{{\it i.e.},\,}
\newcommand{\etal}{{\it et al.\ }}
\newcommand{\eg}{{\it e.g.},\,}
\newcommand{\cf}{{\it cf.\ }}
\newcommand{\vs}{{\it vs.\ }}
\newcommand{\zdot}{\makebox[0pt][l]{.}}
\newcommand{\up}[1]{\ifmmode^{\rm #1}\else$^{\rm #1}$\fi}
\newcommand{\dn}[1]{\ifmmode_{\rm #1}\else$_{\rm #1}$\fi}
\newcommand{\upd}{\up{d}}
\newcommand{\uph}{\up{h}}
\newcommand{\upm}{\up{m}}
\newcommand{\ups}{\up{s}}
\newcommand{\arcd}{\ifmmode^{\circ}\else$^{\circ}$\fi}
\newcommand{\arcm}{\ifmmode{'}\else$'$\fi}
\newcommand{\arcs}{\ifmmode{''}\else$''$\fi}
\newcommand{\MS}{{\rm M}\ifmmode_{\odot}\else$_{\odot}$\fi}
\newcommand{\RS}{{\rm R}\ifmmode_{\odot}\else$_{\odot}$\fi}
\newcommand{\LS}{{\rm L}\ifmmode_{\odot}\else$_{\odot}$\fi}

\newcommand{\Abstract}[2]{{\footnotesize\begin{center}ABSTRACT\end{center}
\vspace{1mm}\par#1\par
\noindent
{~}{\it #2}}}

\newcommand{\TabCap}[2]{\begin{center}\parbox[t]{#1}{\begin{center}
  \vglue-9mm
  \small {\spaceskip 2pt plus 1pt minus 1pt T a b l e}
  \refstepcounter{table}\thetable \\[1mm]
  \footnotesize #2 \end{center}}\end{center}}

\newcommand{\TableSep}[2]{\begin{table}[p]\vspace{#1}
\TabCap{#2}\end{table}}

\newcommand{\FigCap}[1]{\footnotesize\par\noindent Fig.\  %
  \refstepcounter{figure}\thefigure. #1\par}

\newcommand{\TableFont}{\footnotesize}
\newcommand{\TableFontIt}{\ttit}
\newcommand{\SetTableFont}[1]{\renewcommand{\TableFont}{#1}}

\newcommand{\MakeTable}[4]{\begin{table}[htb]\TabCap{#2}{#3}
  \begin{center} \TableFont \begin{tabular}{#1} #4 
  \end{tabular}\end{center}\end{table}}

\newcommand{\MakeTableSep}[4]{\begin{table}[p]\TabCap{#2}{#3}
  \begin{center} \TableFont \begin{tabular}{#1} #4 
  \end{tabular}\end{center}\end{table}}

\newenvironment{references}%
{
\footnotesize \frenchspacing
\renewcommand{\thesection}{}
\renewcommand{\in}{{\rm in }}
\renewcommand{\AA}{Astron.\ Astrophys.}
\newcommand{\AAS}{Astron.~Astrophys.~Suppl.~Ser.}
\newcommand{\ApJ}{Astrophys.\ J.}
\newcommand{\ApJS}{Astrophys.\ J.~Suppl.~Ser.}
\newcommand{\ApJL}{Astrophys.\ J.~Letters}
\newcommand{\AJ}{Astron.\ J.}
\newcommand{\IBVS}{IBVS}
\newcommand{\PASP}{P.A.S.P.}
\newcommand{\Acta}{Acta Astron.}
\newcommand{\MNRAS}{MNRAS}
\renewcommand{\and}{{\rm and }}
\section{{\rm REFERENCES}}
\sloppy \hyphenpenalty10000
\begin{list}{}{\leftmargin1cm\listparindent-1cm
\itemindent\listparindent\parsep0pt\itemsep0pt}}%
{\end{list}\vspace{2mm}}

\def\TYLDA{~}
\newlength{\DW}
\settowidth{\DW}{0}
\newcommand{\dw}{\hspace{\DW}}

\newcommand{\refitem}[5]{\item[]{#1} #2%
\def\REFARG{#3}\ifx\REFARG\TYLDA\else, {\it#3}\fi
\def\REFARG{#4}\ifx\REFARG\TYLDA\else, {\bf#4}\fi
\def\REFARG{#5}\ifx\REFARG\TYLDA\else, {#5}\fi.}

\newcommand{\Section}[1]{\section{#1}}
\newcommand{\Subsection}[1]{\subsection{#1}}
\newcommand{\Acknow}[1]{\par\vspace{5mm}{\bf Acknowledgements.} #1}
\pagestyle{myheadings}

\newfont{\bb}{ptmbi8t at 12pt}
\newcommand{\xrule}{\rule{0pt}{2.5ex}}
\newcommand{\xxrule}{\rule[-1.8ex]{0pt}{4.5ex}}
\def\thefootnote{\fnsymbol{footnote}}
\begin{center}
{\Large\bf The Optical Gravitational Lensing Experiment.\\
\vskip3pt
Ages of about 600 Star Clusters from the LMC\footnote{Based on observations 
obtained with the 1.3~m Warsaw telescope at the Las Campanas Observatory of 
the Carnegie Institution of Washington.}} 

\vskip.5cm
G.~~P~i~e~t~r~z~y~\'n~s~k~i$^{1,2}$~~ and~~A.~~U~d~a~l~s~k~i$^2$
\vskip3mm
$^1$ Universidad de Concepci{\'o}n, Departamento de Fisica,
Casilla 160--C, Concepci{\'o}n, Chile\\
$^2$Warsaw University Observatory, Al.~Ujazdowskie~4, 00-478~Warszawa, Poland\\
e-mail: (pietrzyn,udalski)@sirius.astrouw.edu.pl
\end{center}

\Abstract{We present results of age determination based on the standard 
procedure of isochrone fitting for about 600 star clusters younger than about 
1.2~Gyr from the central parts of the LMC. Comparison of age distributions of 
star clusters from the LMC and the SMC shows that cluster formation histories 
are different in these galaxies. The age distribution of the LMC clusters 
reveals bursty nature of cluster formation in this galaxy, with contrast to 
the relatively uniform distribution of cluster ages in the SMC. We detected 
three extended peaks in the distribution of ages of LMC clusters, located at 
about 7~Myr, 125~Myr and 0.8~Gyr. All detected peaks have complex structure. 
While the structure of the youngest and the oldest peaks may be spurious due 
to accuracy of age determination, in the middle peak two evident sub-peaks at 
100~Myr and 160~Myr are clearly seen. Similar peaks are seen in the 
distribution of ages of clusters from the SMC, which indicates that increased 
cluster formation ratio during these periods might be caused by the last 
encounter between the Magellanic Clouds. The sample of clusters from the LMC 
is overabundant in very young (${<10}$~Myr) and old (${>400}$~Myr) 
clusters, and underabundant in clusters having ages in the range from 80 to 
15~Myr with respect to the studied sample of clusters from the 
SMC.}{Magellanic Clouds -- Galaxies: star clusters} 

\Section{Introduction}
The analysis of rich systems of clusters in the Magellanic Clouds and in the 
Galaxy provides many information about parent galaxies and processes of star 
and cluster formation. Previous investigations revealed that star clusters 
from the Magellanic Clouds and from the Galaxy differ in a number of important 
respects (van den Bergh 1991). In particular, the cluster formation histories 
in these galaxies are very different. In the Milky Way there are two distinct 
classes of old massive globular clusters and less massive, young open ones. In 
the Magellanic Clouds there are also objects of intermediate mass. The age 
distribution of clusters from the LMC shows bursts of star formation and 
quiescent periods in contrast to the continuous age distribution of clusters 
from the SMC (van den Bergh 1991). Unfortunately, the number of cluster from 
the Magellanic Cloud having precise data for age determination was still to 
small to make final statement about star cluster formation history in these 
galaxies. In particular information about the ages of significant fraction of 
the most numerous clusters, younger than one Gyr, was lacking. 

The microlensing experiments have been providing us with incredible amount of 
photometric data. The observations collected in the course of the OGLE-II 
project (Udalski, Kubiak and Szyma{\'n}ski 1997), the only microlensing 
experiment that uses standard {\it BVI} passbands, are especially well suited 
for detecting and analysis of properties of star clusters from the Magellanic 
Clouds. In a series of paper we present results of our investigations. Using 
the automated and algorithmic method the catalogs of almost 1000 clusters from 
the Magellanic Clouds (Pietrzy{\'n}ski \etal 1998, 1999) were constructed. 
Based on these catalogs 99 multiple cluster candidates were selected and 
described by Pietrzy{\'n}ski and Udalski (1999c, 2000). Many observations 
(typically 400) for each star collected during several years provide unique 
possibility to explore populations of variable stars from the Magellanic 
Clouds clusters (Pietrzy{\'n}ski and Udalski 1999 ad). Precise photometry 
down to ${V\approx21.5}$~mag (Udalski \etal 1998) allowed us to derive ages 
for 93 clusters younger than 1~Gyr from the SMC (Pietrzy{\'n}ski and Udalski 
1999b). In this contribution we present the catalog of ages of about 600 
clusters from the Large Magellanic Cloud and compare distributions of ages of 
clusters from the SMC and LMC. 

\Section{Ages of the LMC Clusters}
\vspace*{-9pt}
\Subsection{Age Determination}
\vspace*{-5pt}
Unambiguous distinction between field and cluster stars in the dense regions 
of the LMC is very difficult. We tried to eliminate field star contamination of 
cluster CMD by examining the CMDs of the fields located around a given cluster 
as well as using the procedure of statistical subtraction of field stars 
(Mateo and Hodge 1986). The interstellar reddening was taken from Udalski 
\etal (1999). After dereddening and correcting for field star contamination, 
reliable ages were derived for about 600 relatively reach star clusters using 
standard procedure of isochrone fitting. In some cases two different, equally 
good age determinations are possible because of not perfect field star 
subtraction. 

The isochrones covering the wide range of chemical composition and stellar 
masses (Bertelli \etal 1994) were used. Metallicity of ${Z=0.008}$ and distance 
modulus of 18.24~mag (Udalski 2000) were adopted. The accuracy of our age 
determination depends on age, richness and location of a given cluster. For 
clusters older than about 1.2~Gyr the turn off point is located close to the 
limit of OGLE-II photometry (${V\approx21.5}$~mag) and therefore no reliable 
age determination was possible. For clusters possessing few stars and those 
located in very dense fields a wide variety of isochrones may be fitted. The 
accuracy of our procedure of age determination was estimated as the half of 
the age difference of two marginally fitting isochrones. It is worth noting 
that our results weakly depend on the assumed distance scale. Change in 
the distance modulus by about 0.15 mag produces difference in ages by less 
than a few percents, which is much less than derived errors. 

The results of age determination for about 600 clusters are presented in 
Table~1.

\Subsection{Age Distribution}
Fig.~1 presents distribution of ages of about 700 clusters younger than about 
1.2~Gyr from the Magellanic Clouds located in the OGLE-II fields 
(Pietrzy{\'n}ski and Udalski 1999b, this paper). Contrary to the rather 
uniform distribution of ages of clusters from the SMC, the bursts of cluster 
formation and quiescent periods are clearly visible in the distribution of 
ages of clusters from the LMC: three evident peaks located at about 7~Myr, 
125~Myr and 0.8~Gyr ago are visible. They are separated by periods when 
relatively few clusters were formed. All three peaks show complex structure. 
However, it should be stressed that age determination for objects older than 
about 1~Gyr and younger than 10~Myr is not very accurate and the structure of 
the oldest and the youngest peaks may be spurious. During the period of 
enhanced cluster formation, which lasted from about 100~Myr to 200~Myr, two 
evident, very likely real peaks are seen. They are located at about 100~Myr 
and 160~Myr. In spite of much lower statistics similar peaks are seen in the 
distribution of ages of clusters from the SMC. This fact suggests that the 
enhanced cluster formation rate during these periods may be caused by tidal 
forces induced during the last encounter between the Magellanic Clouds 
(Gardiner \etal 1994). It should be stressed that observed age distribution 
depends not only on the rate of cluster formation but also on the efficiency 
of processes of their disintegration. 
\begin{figure}[p]
\includegraphics[width=10cm]{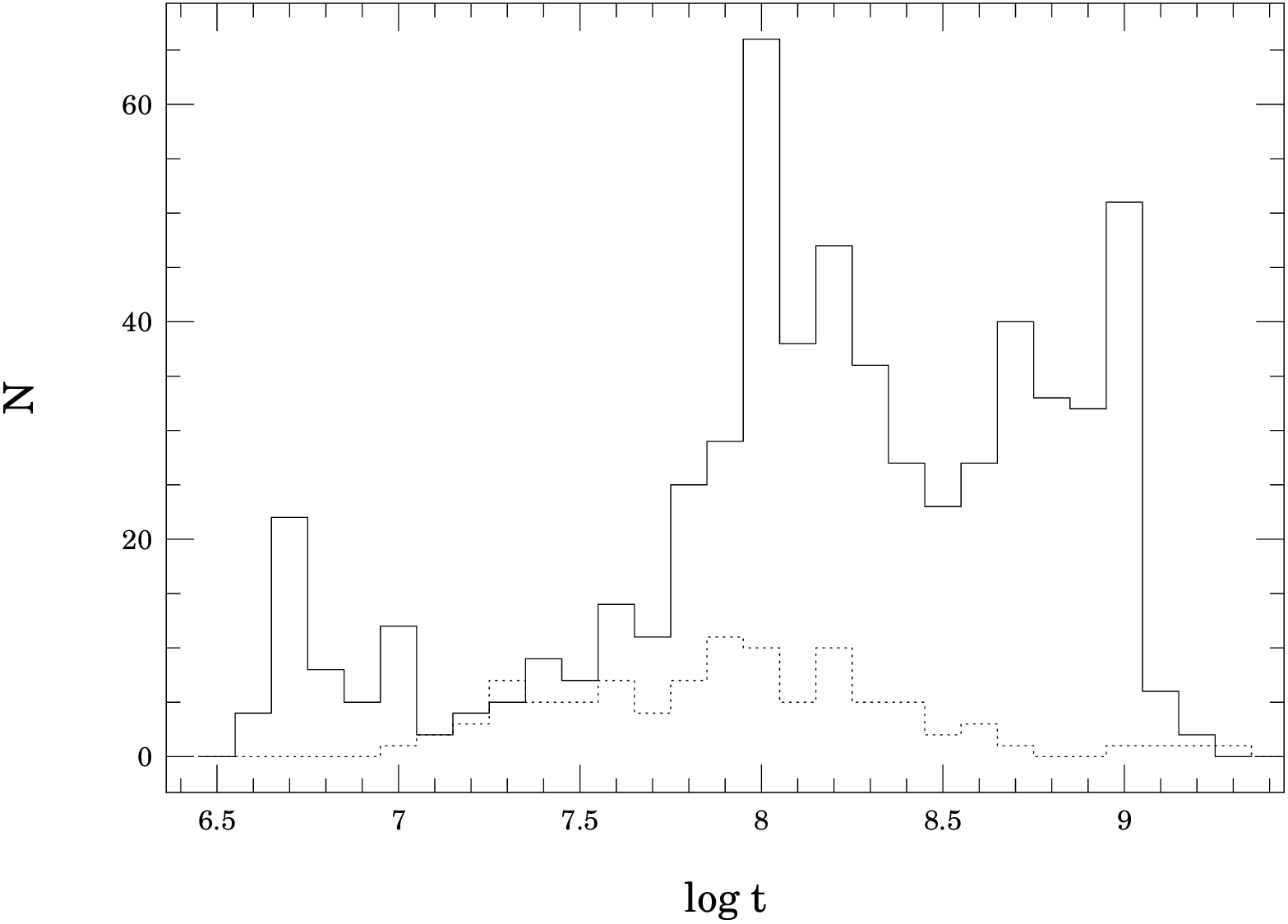}
\FigCap{Age distribution of clusters from the LMC -- solid line, and the SMC 
-- dotted line.}
\vskip1cm
\includegraphics[width=10cm]{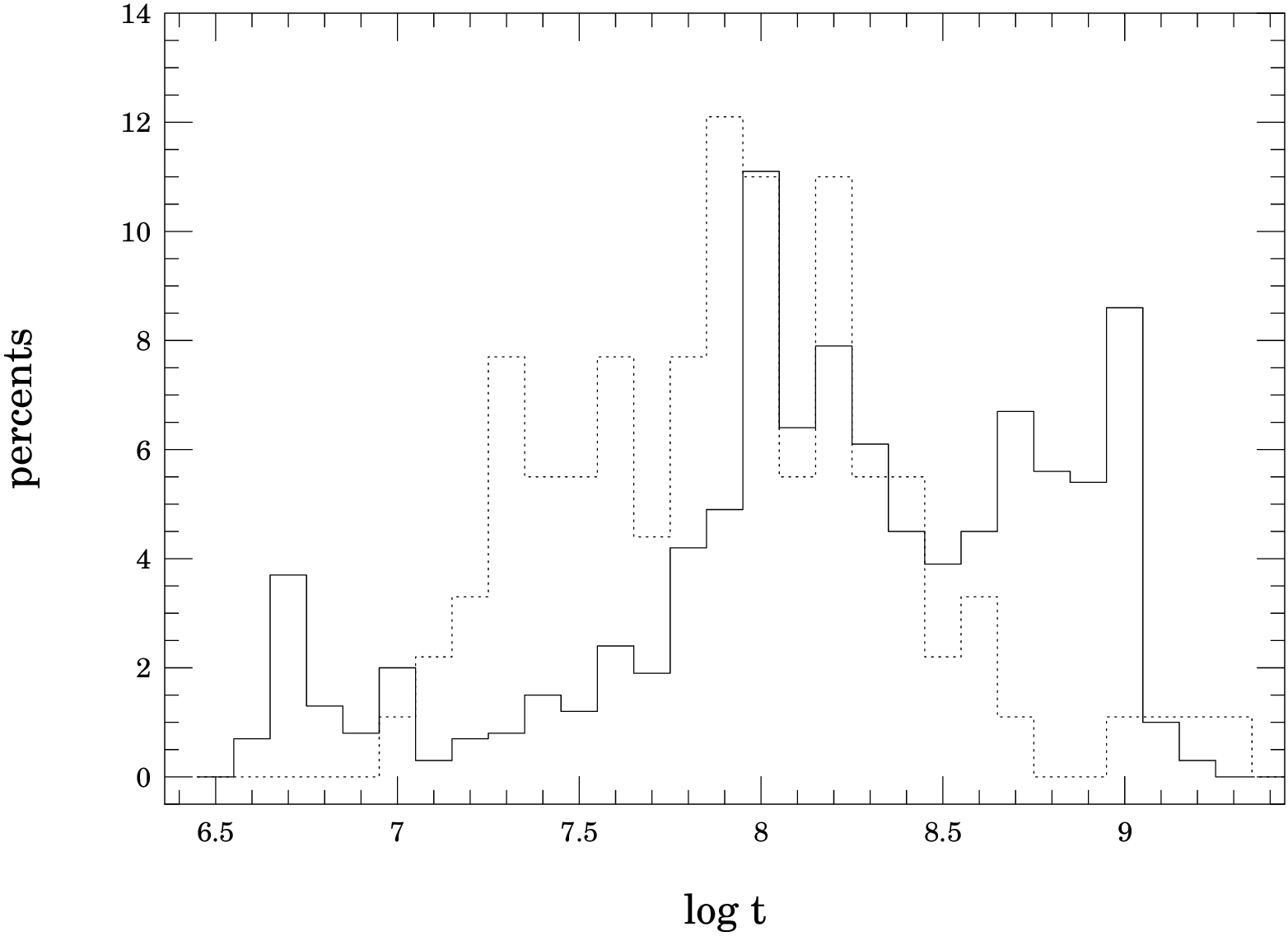}
\FigCap{Normalized age distribution of clusters from the LMC -- solid line, 
and the SMC -- dotted line.}
\end{figure}

In order to make the comparison easier we normalized the age distributions of 
clusters from the LMC and SMC and multiplied by 100, to express them in 
percents (see Fig.~2). The central parts of the LMC possess more young 
(${<10}$~Myr) and older (${>400}$~Myr) clusters than were detected in the main 
body of the SMC. On the other hand the cluster sample from the SMC comprises 
relatively more clusters with ages in the range from 15~Myr to 80~Myr than the 
LMC one. 

More detailed comparison of cluster formation rates in these galaxies requires 
more numerous sample of ages of clusters from the SMC and more precise age 
determination, based on deeper photometry for clusters older than about 1~Gyr. 

\Section{Summary}
We present results of age determination based on the standard procedure of 
isochrone fitting for about 600 star clusters from the central parts of the 
LMC. Based on the dataset, which is order of magnitude larger than available so 
far, we confirmed that cluster formation history of the LMC and the SMC are 
different. The distribution of ages of clusters from the LMC reveals periods of 
bursty cluster formation and quiescent periods. We detected three extended 
periods of enhanced cluster formation at about 7~Myr, 125~Myr and 0.8~Gyr ago. 
No such features are seen in the relatively uniform age distribution of the 
SMC clusters. Both age distributions reveal peaks located at about 100~Myr 
and 160~Myr, which might be connected with the last encounter of the 
Magellanic Clouds. 

\Acknow{The paper was partly supported by the Polish KBN grants 2P03D00814 to 
A.~Udalski. Partial support for the OGLE project was provided with the NSF 
grant AST-9530478 to B.~Paczy\'nski.}

\renewcommand{\arraystretch}{0.9}
\MakeTableSep{|c|c|c|c|c|c|}{12.5cm}{Ages of the LMC clusters}%1
{\hline
Name  &  $\alpha_{2000}$ & $\delta_{2000}$ & Radius
& $\log t$ & $\sigma_{\log t}$\\
OGLE-CL-& & & [\arcs] & & \\
\hline\xrule
LMC0001 & $4\uph59\upm59\zdot\ups63$ & $-69\arcd30\arcm39\zdot\arcs2$&   13  & 8.33 & 0.05 \\ 
LMC0002 & $5\uph00\upm10\zdot\ups93$ & $-68\arcd36\arcm54\zdot\arcs8$&   10  & 8.9 & 0.1 \\ 
LMC0003 & $5\uph00\upm14\zdot\ups53$ & $-69\arcd09\arcm22\zdot\arcs3$&   18  & 8.7 & 0.1 \\ 
LMC0004 & $5\uph00\upm21\zdot\ups05$ & $-69\arcd06\arcm25\zdot\arcs0$&   16  & 7.6,8.5 & 0.01 \\ 
LMC0005 & $5\uph00\upm26\zdot\ups85$ & $-68\arcd46\arcm22\zdot\arcs2$&   16  & 8.15 & 0.05 \\ 
LMC0006 & $5\uph00\upm28\zdot\ups96$ & $-68\arcd38\arcm27\zdot\arcs5$&   13  & 8.85 & 0.05 \\ 
LMC0007 & $5\uph00\upm41\zdot\ups13$ & $-69\arcd20\arcm28\zdot\arcs9$&   17  & 9.05 & 0.05 \\ 
LMC0008 & $5\uph01\upm04\zdot\ups50$ & $-69\arcd05\arcm03\zdot\arcs3$&   23  & 7.93 & 0.05 \\ 
LMC0009 & $5\uph01\upm14\zdot\ups91$ & $-67\arcd32\arcm20\zdot\arcs0$&   24  & 8.1 & 0.05 \\ 
LMC0010 & $5\uph01\upm16\zdot\ups43$ & $-69\arcd12\arcm01\zdot\arcs3$&   18  & 8.0 & 0.1 \\ 
LMC0012 & $5\uph01\upm22\zdot\ups46$ & $-67\arcd17\arcm41\zdot\arcs2$&   20  & 8.7 & 0.1 \\ 
LMC0013 & $5\uph01\upm23\zdot\ups79$ & $-68\arcd52\arcm22\zdot\arcs5$&   20  & 8.7 & 0.05 \\ 
LMC0015 & $5\uph01\upm29\zdot\ups20$ & $-68\arcd42\arcm43\zdot\arcs8$&   26  & 8.25 & 0.08 \\ 
LMC0016 & $5\uph01\upm29\zdot\ups29$ & $-67\arcd20\arcm59\zdot\arcs2$&   10  & 8.5 & 0.1 \\ 
LMC0017 & $5\uph01\upm31\zdot\ups83$ & $-69\arcd32\arcm05\zdot\arcs5$&   20  & 7.4 & 0.1 \\ 
LMC0018 & $5\uph01\upm36\zdot\ups95$ & $-69\arcd02\arcm17\zdot\arcs6$&   30  & 9.0 & 0.1 \\ 
LMC0020 & $5\uph01\upm45\zdot\ups39$ & $-67\arcd34\arcm00\zdot\arcs7$&    8  & 7.9 & 0.2 \\ 
LMC0021 & $5\uph01\upm45\zdot\ups47$ & $-67\arcd05\arcm43\zdot\arcs9$&   18  & 8.15 & 0.1 \\ 
LMC0022 & $5\uph01\upm46\zdot\ups27$ & $-69\arcd23\arcm56\zdot\arcs6$&    8  & 9 & 0.1 \\ 
LMC0023 & $5\uph01\upm48\zdot\ups47$ & $-67\arcd28\arcm24\zdot\arcs5$&   24  & 7.95 & 0.05 \\ 
LMC0025 & $5\uph01\upm51\zdot\ups69$ & $-69\arcd12\arcm51\zdot\arcs5$&   32  & 8.6 & 0.1 \\ 
LMC0028 & $5\uph02\upm00\zdot\ups52$ & $-67\arcd22\arcm07\zdot\arcs1$&   47  & 8.1 & 0.05 \\ 
LMC0029 & $5\uph02\upm06\zdot\ups46$ & $-66\arcd45\arcm19\zdot\arcs4$&   12  & 8.7 & 0.1 \\ 
LMC0030 & $5\uph02\upm18\zdot\ups92$ & $-69\arcd32\arcm05\zdot\arcs8$&   17  & 8.7 & 0.05 \\ 
LMC0031 & $5\uph02\upm33\zdot\ups05$ & $-68\arcd49\arcm21\zdot\arcs6$&   17  & 7.9 & 0.1 \\ 
LMC0032 & $5\uph02\upm41\zdot\ups00$ & $-69\arcd31\arcm36\zdot\arcs9$&    8  & 8.7 & 0.1 \\ 
LMC0033 & $5\uph02\upm54\zdot\ups78$ & $-68\arcd46\arcm16\zdot\arcs0$&   22  & 7.8 & 0.1 \\ 
LMC0034 & $5\uph03\upm04\zdot\ups51$ & $-69\arcd28\arcm10\zdot\arcs4$&   20  & 9.0 & 0.1 \\ 
LMC0035 & $5\uph03\upm05\zdot\ups95$ & $-69\arcd02\arcm14\zdot\arcs9$&   25  & 7.9 & 0.05 \\ 
LMC0037 & $5\uph03\upm19\zdot\ups52$ & $-66\arcd58\arcm51\zdot\arcs9$&   18  & $<$6.7 & -- \\ 
LMC0038 & $5\uph03\upm24\zdot\ups16$ & $-68\arcd51\arcm29\zdot\arcs7$&   16  & 7.85 & 0.05 \\ 
LMC0039 & $5\uph03\upm26\zdot\ups51$ & $-67\arcd15\arcm12\zdot\arcs5$&   19  & $<$6.7 & -- \\ 
LMC0041 & $5\uph03\upm33\zdot\ups69$ & $-67\arcd37\arcm33\zdot\arcs9$&   18  & 8.9 & 0.08 \\ 
LMC0042 & $5\uph03\upm38\zdot\ups63$ & $-69\arcd23\arcm10\zdot\arcs4$&   23  & 8.9 & 0.05 \\ 
LMC0043 & $5\uph03\upm38\zdot\ups64$ & $-68\arcd58\arcm44\zdot\arcs3$&   13  & 8.5 & 0.05 \\ 
LMC0044 & $5\uph03\upm42\zdot\ups13$ & $-68\arcd58\arcm06\zdot\arcs3$&   14  & 7.7 & 0.1 \\ 
LMC0045 & $5\uph03\upm44\zdot\ups77$ & $-66\arcd43\arcm33\zdot\arcs9$&    9  & 7.8 & 0.1 \\ 
LMC0046 & $5\uph03\upm46\zdot\ups14$ & $-67\arcd18\arcm15\zdot\arcs0$&   21  & $<$6.8 & -- \\ 
LMC0047 & $5\uph03\upm47\zdot\ups88$ & $-68\arcd51\arcm01\zdot\arcs6$&   19  & 7.9 & 0.1 \\ 
LMC0049 & $5\uph03\upm51\zdot\ups68$ & $-67\arcd15\arcm40\zdot\arcs9$&   19  & 6.9 & 0.2 \\ 
LMC0050 & $5\uph03\upm58\zdot\ups11$ & $-67\arcd24\arcm02\zdot\arcs5$&   10  & 8.8 & 0.05 \\ 
LMC0052 & $5\uph04\upm11\zdot\ups79$ & $-69\arcd18\arcm10\zdot\arcs4$&   19  & 8.1 & 0.1 \\ 
LMC0053 & $5\uph04\upm19\zdot\ups30$ & $-69\arcd21\arcm23\zdot\arcs2$&   10  & 8.8 & 0.1 \\ 
LMC0054 & $5\uph04\upm19\zdot\ups43$ & $-68\arcd55\arcm39\zdot\arcs0$&   19  & 7.6 & 0.1 \\ 
LMC0055 & $5\uph04\upm21\zdot\ups39$ & $-69\arcd23\arcm16\zdot\arcs2$&   20  & 8.2 & 0.1 \\ 
LMC0056 & $5\uph04\upm23\zdot\ups91$ & $-69\arcd28\arcm01\zdot\arcs7$&   26  & 8.55 & 0.1 \\ 
LMC0057 & $5\uph04\upm24\zdot\ups94$ & $-69\arcd20\arcm59\zdot\arcs7$&   13  & 8.85 & 0.05 \\ 
LMC0058 & $5\uph04\upm30\zdot\ups43$ & $-69\arcd09\arcm21\zdot\arcs8$&   21  & 8.4 & 0.1 \\ 
LMC0059 & $5\uph04\upm30\zdot\ups57$ & $-69\arcd21\arcm18\zdot\arcs3$&   20  & 8.8 & 0.1 \\ 
LMC0060 & $5\uph04\upm33\zdot\ups73$ & $-69\arcd01\arcm02\zdot\arcs9$&   18  & 8.8 & 0.1 \\ 
\hline}

\setcounter{table}{0}
\MakeTableSep{|c|c|c|c|c|c|}{12.5cm}{continued}%2
{\hline
Name  &  $\alpha_{2000}$ & $\delta_{2000}$ & Radius
& $\log t$ & $\sigma_{\log t}$\\
OGLE-CL-& & & [\arcs] & & \\
\hline\xrule
LMC0061 & $5\uph04\upm39\zdot\ups09$ & $-69\arcd20\arcm26\zdot\arcs1$&   25  & 8.4 & 0.1 \\ 
LMC0062 & $5\uph04\upm41\zdot\ups35$ & $-69\arcd14\arcm50\zdot\arcs6$&   18  & 8.7 & 0.1 \\ 
LMC0063 & $5\uph04\upm44\zdot\ups89$ & $-68\arcd59\arcm03\zdot\arcs8$&   19  & 8.4 & 0.1 \\ 
LMC0064 & $5\uph04\upm50\zdot\ups43$ & $-68\arcd59\arcm16\zdot\arcs2$&   20  & 8.4 & 0.05 \\ 
LMC0065 & $5\uph04\upm57\zdot\ups23$ & $-70\arcd01\arcm08\zdot\arcs4$&   18  & 7.3 & 0.1 \\ 
LMC0066 & $5\uph05\upm00\zdot\ups64$ & $-68\arcd45\arcm01\zdot\arcs3$&   14  & 8.0 & 0.05 \\ 
LMC0067 & $5\uph05\upm02\zdot\ups99$ & $-68\arcd54\arcm37\zdot\arcs4$&   31  & 8.85 & 0.15 \\ 
LMC0068 & $5\uph05\upm06\zdot\ups14$ & $-69\arcd03\arcm12\zdot\arcs3$&   18  & 8 & 0.1 \\ 
LMC0069 & $5\uph05\upm07\zdot\ups05$ & $-69\arcd24\arcm14\zdot\arcs3$&   70  & $>$9.2 & -- \\ 
LMC0071 & $5\uph05\upm09\zdot\ups72$ & $-68\arcd57\arcm23\zdot\arcs8$&   15  & 7.8 & 0.1 \\ 
LMC0072 & $5\uph05\upm12\zdot\ups13$ & $-69\arcd12\arcm26\zdot\arcs5$&   23  & 7.9 & 0.15 \\ 
LMC0073 & $5\uph05\upm12\zdot\ups14$ & $-68\arcd33\arcm10\zdot\arcs2$&   19  & 8.5 & 0.1 \\ 
LMC0074 & $5\uph05\upm13\zdot\ups74$ & $-69\arcd22\arcm11\zdot\arcs9$&   14  & 8.65 & 0.1 \\ 
LMC0075 & $5\uph05\upm14\zdot\ups14$ & $-68\arcd44\arcm34\zdot\arcs4$&   18  & 7.9 & 0.05 \\ 
LMC0076 & $5\uph05\upm14\zdot\ups36$ & $-66\arcd42\arcm06\zdot\arcs9$&    9  & $<$7.0 & -- \\ 
LMC0077 & $5\uph05\upm18\zdot\ups53$ & $-68\arcd43\arcm33\zdot\arcs7$&   19  & 7.9 & 0.1 \\ 
LMC0078 & $5\uph05\upm19\zdot\ups18$ & $-68\arcd44\arcm14\zdot\arcs7$&   30  & 8.0 & 0.1 \\ 
LMC0079 & $5\uph05\upm23\zdot\ups83$ & $-69\arcd20\arcm23\zdot\arcs8$&   16  & 8.8 & 0.15 \\ 
LMC0080 & $5\uph05\upm24\zdot\ups96$ & $-68\arcd30\arcm01\zdot\arcs9$&   18  & 7.4 & 0.05 \\ 
LMC0081 & $5\uph05\upm35\zdot\ups79$ & $-68\arcd37\arcm42\zdot\arcs5$&   35  & 8.5 & 0.05 \\ 
LMC0082 & $5\uph05\upm36\zdot\ups94$ & $-68\arcd43\arcm06\zdot\arcs4$&   25  & 7.8 & 0.1 \\ 
LMC0083 & $5\uph05\upm40\zdot\ups09$ & $-68\arcd38\arcm11\zdot\arcs9$&   25  & 7.8 & 0.05 \\ 
LMC0084 & $5\uph05\upm42\zdot\ups43$ & $-69\arcd02\arcm20\zdot\arcs5$&    8  & 8.9 & 0.1 \\ 
LMC0085 & $5\uph05\upm44\zdot\ups44$ & $-68\arcd30\arcm24\zdot\arcs6$&   14  & 8.9 & 0.1 \\ 
LMC0088 & $5\uph05\upm54\zdot\ups61$ & $-66\arcd43\arcm57\zdot\arcs7$&   16  & 8.2 & 0.15 \\ 
LMC0089 & $5\uph05\upm55\zdot\ups36$ & $-68\arcd57\arcm04\zdot\arcs8$&   17  & 8.0 & 0.1 \\ 
LMC0090 & $5\uph05\upm55\zdot\ups63$ & $-68\arcd37\arcm42\zdot\arcs8$&   27  & 7.8 & 0.1 \\ 
LMC0091 & $5\uph06\upm00\zdot\ups72$ & $-68\arcd33\arcm21\zdot\arcs0$&   18  & 8.9 & 0.1 \\ 
LMC0092 & $5\uph06\upm02\zdot\ups27$ & $-68\arcd57\arcm22\zdot\arcs2$&   14  & 7.2 & 0.2 \\ 
LMC0093 & $5\uph06\upm02\zdot\ups89$ & $-68\arcd37\arcm41\zdot\arcs6$&   25  & 8.0 & 0.05 \\ 
LMC0094 & $5\uph06\upm06\zdot\ups18$ & $-67\arcd02\arcm02\zdot\arcs8$&   16  & $<$6.9 & -- \\ 
LMC0095 & $5\uph06\upm06\zdot\ups33$ & $-68\arcd22\arcm00\zdot\arcs9$&   12  & 8.7 & 0.1 \\ 
LMC0096 & $5\uph06\upm07\zdot\ups30$ & $-69\arcd11\arcm03\zdot\arcs7$&   17  & 8.5 & 0.1 \\ 
LMC0097 & $5\uph06\upm08\zdot\ups81$ & $-68\arcd26\arcm45\zdot\arcs2$&   25  & 8.0 & 0.05 \\ 
LMC0098 & $5\uph06\upm10\zdot\ups08$ & $-70\arcd00\arcm52\zdot\arcs2$&   22  & 8.1 & 0.05 \\ 
LMC0099 & $5\uph06\upm11\zdot\ups55$ & $-69\arcd58\arcm21\zdot\arcs9$&   19  & 8.2 & 0.1 \\ 
LMC0100 & $5\uph06\upm12\zdot\ups42$ & $-69\arcd03\arcm26\zdot\arcs5$&   20  & 6.7 & 0.1 \\ 
LMC0101 & $5\uph06\upm12\zdot\ups80$ & $-68\arcd50\arcm54\zdot\arcs0$&   18  & 8.5 & 0.1 \\ 
LMC0102 & $5\uph06\upm22\zdot\ups30$ & $-69\arcd28\arcm04\zdot\arcs7$&   23  & 8.7 & 0.1 \\ 
LMC0103 & $5\uph06\upm24\zdot\ups14$ & $-69\arcd34\arcm06\zdot\arcs1$&   23  & 8.7 & 0.1 \\ 
LMC0104 & $5\uph06\upm24\zdot\ups53$ & $-68\arcd42\arcm18\zdot\arcs8$&   29  & 9.0 & 0.1 \\ 
LMC0105 & $5\uph06\upm24\zdot\ups81$ & $-68\arcd22\arcm29\zdot\arcs5$&   20  & 8.35 & 0.1 \\ 
LMC0106 & $5\uph06\upm27\zdot\ups95$ & $-66\arcd54\arcm21\zdot\arcs4$&   29  & 8.15 & 0.1 \\ 
LMC0107 & $5\uph06\upm33\zdot\ups57$ & $-68\arcd21\arcm47\zdot\arcs3$&   19  & 7.3 & 0.05 \\ 
LMC0109 & $5\uph06\upm34\zdot\ups11$ & $-68\arcd25\arcm38\zdot\arcs2$&   14  & $<$6.7 & -- \\ 
LMC0111 & $5\uph06\upm47\zdot\ups40$ & $-68\arcd36\arcm59\zdot\arcs4$&   41  & 8.0 & 0.1 \\ 
LMC0112 & $5\uph06\upm51\zdot\ups75$ & $-68\arcd52\arcm51\zdot\arcs8$&   24  & 8.7 & 0.1 \\ 
LMC0113 & $5\uph06\upm54\zdot\ups55$ & $-68\arcd43\arcm07\zdot\arcs8$&   27  & 7.9 & 0.1 \\ 
LMC0114 & $5\uph06\upm55\zdot\ups72$ & $-69\arcd25\arcm48\zdot\arcs2$&   12  & 9.05 & 0.1 \\ 
LMC0115 & $5\uph06\upm57\zdot\ups14$ & $-68\arcd39\arcm27\zdot\arcs0$&   32  & 7.9 & 0.1 \\ 
\hline}

\setcounter{table}{0}
\MakeTableSep{|c|c|c|c|c|c|}{12.5cm}{continued}%3
{\hline
Name  &  $\alpha_{2000}$ & $\delta_{2000}$ & Radius
& $\log t$ & $\sigma_{\log t}$\\
OGLE-CL-& & & [\arcs] & & \\
\hline\xrule
LMC0116 & $5\uph06\upm58\zdot\ups34$ & $-69\arcd08\arcm50\zdot\arcs5$&   30  & 7.58 & 0.05 \\ 
LMC0117 & $5\uph06\upm59\zdot\ups17$ & $-69\arcd19\arcm11\zdot\arcs7$&   20  & 8.6 & 0.1 \\ 
LMC0118 & $5\uph07\upm08\zdot\ups54$ & $-68\arcd58\arcm22\zdot\arcs7$&   34  & 7.7 & 0.05 \\ 
LMC0119 & $5\uph07\upm10\zdot\ups92$ & $-68\arcd18\arcm03\zdot\arcs5$&   13  & 7.3 & 0.15 \\ 
LMC0120 & $5\uph07\upm11\zdot\ups38$ & $-69\arcd07\arcm13\zdot\arcs1$&   20  & 8.0 & 0.1 \\ 
LMC0121 & $5\uph07\upm18\zdot\ups75$ & $-67\arcd16\arcm20\zdot\arcs4$&   21  & 8.5 & 0.1 \\ 
LMC0123 & $5\uph07\upm20\zdot\ups75$ & $-66\arcd49\arcm44\zdot\arcs7$&    9  & 8.2 & 0.05 \\ 
LMC0124 & $5\uph07\upm28\zdot\ups07$ & $-68\arcd58\arcm31\zdot\arcs9$&   25  & 8.8 & 0.1 \\ 
LMC0125 & $5\uph07\upm29\zdot\ups76$ & $-68\arcd53\arcm20\zdot\arcs5$&   25  & 9.0 & 0.1 \\ 
LMC0126 & $5\uph07\upm30\zdot\ups27$ & $-67\arcd19\arcm26\zdot\arcs3$&   26  & 7.8 & 0.1 \\ 
LMC0127 & $5\uph07\upm31\zdot\ups93$ & $-67\arcd34\arcm12\zdot\arcs5$&   12  & 7.2 & 0.3 \\ 
LMC0128 & $5\uph07\upm35\zdot\ups25$ & $-67\arcd27\arcm38\zdot\arcs9$&   82  & $>$9.0 & -- \\ 
LMC0129 & $5\uph07\upm38\zdot\ups63$ & $-68\arcd47\arcm45\zdot\arcs9$&   20  & 7.9 & 0.05 \\ 
LMC0132 & $5\uph07\upm51\zdot\ups03$ & $-69\arcd26\arcm11\zdot\arcs0$&   22  & 8.8 & 0.1 \\ 
LMC0133 & $5\uph07\upm55\zdot\ups46$ & $-69\arcd17\arcm57\zdot\arcs3$&   11  & 8.3 & 0.1 \\ 
LMC0134 & $5\uph07\upm55\zdot\ups81$ & $-67\arcd21\arcm28\zdot\arcs3$&   16  & 9 & 0.1 \\ 
LMC0136 & $5\uph08\upm06\zdot\ups57$ & $-69\arcd16\arcm04\zdot\arcs3$&   31  & 8.2 & 0.05 \\ 
LMC0137 & $5\uph08\upm11\zdot\ups34$ & $-69\arcd02\arcm22\zdot\arcs9$&   24  & 8.8 & 0.05 \\ 
LMC0139 & $5\uph08\upm27\zdot\ups66$ & $-66\arcd46\arcm13\zdot\arcs7$&   20  & 8.0 & 0.1 \\ 
LMC0140 & $5\uph08\upm34\zdot\ups99$ & $-69\arcd10\arcm36\zdot\arcs1$&   20  & 8.85 & 0.05 \\ 
LMC0141 & $5\uph08\upm43\zdot\ups59$ & $-69\arcd10\arcm58\zdot\arcs6$&   18  & 7.7,8.2 & 0.1 \\ 
LMC0142 & $5\uph08\upm45\zdot\ups79$ & $-68\arcd45\arcm38\zdot\arcs6$&   63  & 7.9 & 0.05 \\ 
LMC0143 & $5\uph08\upm45\zdot\ups90$ & $-68\arcd41\arcm57\zdot\arcs8$&   11  & 8.55 & 0.05 \\ 
LMC0144 & $5\uph08\upm53\zdot\ups74$ & $-66\arcd47\arcm07\zdot\arcs5$&    9  & $<$7.0 & -- \\ 
LMC0145 & $5\uph08\upm54\zdot\ups55$ & $-68\arcd45\arcm13\zdot\arcs9$&   29  & 8.0 & 0.1 \\ 
LMC0146 & $5\uph08\upm56\zdot\ups85$ & $-69\arcd36\arcm31\zdot\arcs3$&   25  & 9.0 & 0.1 \\ 
LMC0149 & $5\uph09\upm12\zdot\ups95$ & $-69\arcd17\arcm00\zdot\arcs0$&   11  & 9.05 & 0.05 \\ 
LMC0150 & $5\uph09\upm13\zdot\ups36$ & $-69\arcd06\arcm52\zdot\arcs0$&   18  & 8.65 & 0.05 \\ 
LMC0152 & $5\uph09\upm14\zdot\ups74$ & $-68\arcd44\arcm02\zdot\arcs1$&   20  & 8.6 & 0.1 \\ 
LMC0153 & $5\uph09\upm14\zdot\ups75$ & $-69\arcd35\arcm17\zdot\arcs4$&   32  & $>$9 & -- \\ 
LMC0154 & $5\uph09\upm20\zdot\ups10$ & $-68\arcd50\arcm52\zdot\arcs8$&   43  & 7.85 & 0.05 \\ 
LMC0155 & $5\uph09\upm24\zdot\ups99$ & $-68\arcd51\arcm47\zdot\arcs4$&   25  & 7.6 & 0.05 \\ 
LMC0156 & $5\uph09\upm28\zdot\ups43$ & $-68\arcd51\arcm01\zdot\arcs5$&   21  & 7.8 & 0.05 \\ 
LMC0157 & $5\uph09\upm30\zdot\ups39$ & $-69\arcd07\arcm45\zdot\arcs4$&   59  & 8.4 & 0.1 \\ 
LMC0158 & $5\uph09\upm40\zdot\ups19$ & $-69\arcd12\arcm39\zdot\arcs4$&   25  & 8.7 & 0.1 \\ 
LMC0159 & $5\uph09\upm42\zdot\ups27$ & $-69\arcd11\arcm08\zdot\arcs8$&   30  & 7.1 & 0.1 \\ 
LMC0160 & $5\uph09\upm42\zdot\ups92$ & $-68\arcd48\arcm06\zdot\arcs5$&   18  & 8.2 & 0.1 \\ 
LMC0161 & $5\uph09\upm45\zdot\ups66$ & $-68\arcd47\arcm18\zdot\arcs1$&   22  & 8.3 & 0.1 \\ 
LMC0162 & $5\uph09\upm49\zdot\ups23$ & $-69\arcd05\arcm04\zdot\arcs0$&   22  & 8.8 & 0.1 \\ 
LMC0163 & $5\uph09\upm55\zdot\ups81$ & $-69\arcd19\arcm50\zdot\arcs3$&    9  & 7.7 & 0.2 \\ 
LMC0164 & $5\uph09\upm56\zdot\ups09$ & $-68\arcd54\arcm06\zdot\arcs2$&  119  & $<$6.7 & -- \\ 
LMC0165 & $5\uph09\upm57\zdot\ups09$ & $-68\arcd43\arcm55\zdot\arcs8$&   12  & 8.6 & 0.05 \\ 
LMC0167 & $5\uph09\upm59\zdot\ups75$ & $-69\arcd21\arcm14\zdot\arcs4$&   20  & 8.7 & 0.08 \\ 
LMC0168 & $5\uph10\upm02\zdot\ups77$ & $-68\arcd50\arcm00\zdot\arcs5$&   25  & 7.9,8.4 & 0.1 \\ 
LMC0170 & $5\uph10\upm11\zdot\ups15$ & $-69\arcd05\arcm15\zdot\arcs0$&   12  & 8.0 & 0.1 \\ 
LMC0171 & $5\uph10\upm13\zdot\ups42$ & $-68\arcd42\arcm30\zdot\arcs0$&   24  & 8.75 & 0.1 \\ 
LMC0172 & $5\uph10\upm16\zdot\ups22$ & $-69\arcd20\arcm28\zdot\arcs9$&   27  & 7.4 & 0.1 \\ 
LMC0173 & $5\uph10\upm18\zdot\ups54$ & $-69\arcd04\arcm46\zdot\arcs5$&   10  & 7.7 & 0.1 \\ 
LMC0174 & $5\uph10\upm18\zdot\ups79$ & $-69\arcd16\arcm21\zdot\arcs3$&    9  & 8.3 & 0.05 \\ 
LMC0176 & $5\uph10\upm20\zdot\ups23$ & $-68\arcd52\arcm37\zdot\arcs6$&   19  & 8.3 & 0.1 \\ 
\hline}

\setcounter{table}{0}
\MakeTableSep{|c|c|c|c|c|c|}{12.5cm}{continued}%4
{\hline
Name  &  $\alpha_{2000}$ & $\delta_{2000}$ & Radius
& $\log t$ & $\sigma_{\log t}$\\
OGLE-CL-& & & [\arcs] & & \\
\hline\xrule
LMC0177 & $5\uph10\upm22\zdot\ups56$ & $-68\arcd55\arcm41\zdot\arcs8$&   34  & 8.2 & 0.05 \\ 
LMC0178 & $5\uph10\upm27\zdot\ups87$ & $-68\arcd41\arcm54\zdot\arcs8$&   12  & 7.9 & 0.1 \\ 
LMC0179 & $5\uph10\upm29\zdot\ups73$ & $-68\arcd52\arcm21\zdot\arcs5$&   21  & 8.35 & 0.1 \\ 
LMC0180 & $5\uph10\upm30\zdot\ups90$ & $-68\arcd56\arcm03\zdot\arcs1$&   27  & 8.0 & 0.2 \\ 
LMC0182 & $5\uph10\upm32\zdot\ups11$ & $-66\arcd56\arcm24\zdot\arcs1$&   10  & 7.5 & 0.1 \\ 
LMC0183 & $5\uph10\upm32\zdot\ups91$ & $-67\arcd07\arcm38\zdot\arcs7$&   23  & 7.6 & 0.2 \\ 
LMC0184 & $5\uph10\upm35\zdot\ups82$ & $-69\arcd08\arcm47\zdot\arcs8$&   12  & 6.7 & 0.1 \\ 
LMC0185 & $5\uph10\upm39\zdot\ups07$ & $-69\arcd02\arcm31\zdot\arcs0$&   22  & 7.45 & 0.05 \\ 
LMC0186 & $5\uph10\upm39\zdot\ups29$ & $-66\arcd43\arcm44\zdot\arcs7$&    8  & 7.7 & 0.2 \\ 
LMC0187 & $5\uph10\upm39\zdot\ups87$ & $-68\arcd45\arcm13\zdot\arcs0$&   28  & 8.4 & 0.1 \\ 
LMC0188 & $5\uph10\upm40\zdot\ups11$ & $-69\arcd16\arcm26\zdot\arcs5$&   14  & 8.5 & 0.1 \\ 
LMC0189 & $5\uph10\upm42\zdot\ups05$ & $-69\arcd34\arcm34\zdot\arcs4$&   23  & 8.8 & 0.05 \\ 
LMC0190 & $5\uph10\upm43\zdot\ups63$ & $-67\arcd04\arcm49\zdot\arcs2$&   10  & 7.0 & 0.2 \\ 
LMC0191 & $5\uph10\upm53\zdot\ups54$ & $-67\arcd28\arcm16\zdot\arcs1$&   12  & 8.65 & 0.1 \\ 
LMC0192 & $5\uph10\upm55\zdot\ups82$ & $-69\arcd33\arcm33\zdot\arcs5$&   25  & 8.65 & 0.05 \\ 
LMC0193 & $5\uph10\upm55\zdot\ups91$ & $-68\arcd56\arcm36\zdot\arcs3$&   12  & 8.75 & 0.05 \\ 
LMC0194 & $5\uph10\upm56\zdot\ups04$ & $-67\arcd37\arcm36\zdot\arcs0$&   16  & 8.7 & 0.1 \\ 
LMC0196 & $5\uph10\upm59\zdot\ups82$ & $-66\arcd44\arcm30\zdot\arcs5$&   27  & 7.6 & 0.15 \\ 
LMC0198 & $5\uph11\upm06\zdot\ups17$ & $-69\arcd10\arcm19\zdot\arcs5$&   13  & 7.4 & 0.2 \\ 
LMC0200 & $5\uph11\upm21\zdot\ups50$ & $-69\arcd18\arcm37\zdot\arcs6$&   12  & 7.95 & 0.1 \\ 
LMC0201 & $5\uph11\upm27\zdot\ups91$ & $-68\arcd53\arcm45\zdot\arcs9$&   15  & 8.2 & 0.05 \\ 
LMC0202 & $5\uph11\upm28\zdot\ups28$ & $-68\arcd51\arcm03\zdot\arcs6$&    9  & 8.0 & 0.1 \\ 
LMC0203 & $5\uph11\upm31\zdot\ups45$ & $-66\arcd58\arcm32\zdot\arcs2$&   37  & 8.9 & 0.1 \\ 
LMC0205 & $5\uph11\upm34\zdot\ups35$ & $-69\arcd06\arcm34\zdot\arcs1$&   32  & 8.2 & 0.15 \\ 
LMC0206 & $5\uph11\upm40\zdot\ups14$ & $-68\arcd43\arcm35\zdot\arcs9$&   30  & 7.85 & 0.1 \\ 
LMC0207 & $5\uph11\upm40\zdot\ups85$ & $-67\arcd33\arcm56\zdot\arcs3$&   27  & 8.8 & 0.1 \\ 
LMC0208 & $5\uph11\upm43\zdot\ups74$ & $-68\arcd47\arcm08\zdot\arcs9$&   13  & 8.3 & 0.1 \\ 
LMC0209 & $5\uph12\upm00\zdot\ups99$ & $-69\arcd12\arcm04\zdot\arcs4$&   39  & 8.2 & 0.1 \\ 
LMC0210 & $5\uph12\upm03\zdot\ups06$ & $-69\arcd17\arcm11\zdot\arcs8$&   18  & 7.9 & 0.1 \\ 
LMC0211 & $5\uph12\upm03\zdot\ups79$ & $-69\arcd12\arcm53\zdot\arcs5$&   26  & 8.2 & 0.05 \\ 
LMC0212 & $5\uph12\upm08\zdot\ups79$ & $-69\arcd16\arcm44\zdot\arcs5$&   20  & 7.9 & 0.1 \\ 
LMC0213 & $5\uph12\upm09\zdot\ups46$ & $-68\arcd54\arcm44\zdot\arcs3$&   12  & 8.9 & 0.05 \\ 
LMC0214 & $5\uph12\upm13\zdot\ups20$ & $-68\arcd57\arcm04\zdot\arcs5$&   18  & 8.4 & 0.1 \\ 
LMC0215 & $5\uph12\upm14\zdot\ups91$ & $-68\arcd55\arcm52\zdot\arcs1$&   18  & 8.2 & 0.08 \\ 
LMC0216 & $5\uph12\upm14\zdot\ups92$ & $-69\arcd25\arcm03\zdot\arcs6$&   23  & 8.9 & 0.1 \\ 
LMC0217 & $5\uph12\upm15\zdot\ups45$ & $-67\arcd04\arcm25\zdot\arcs8$&   11  & 8.65 & 0.1 \\ 
LMC0218 & $5\uph12\upm17\zdot\ups18$ & $-69\arcd17\arcm31\zdot\arcs9$&   13  & 9.1 & 0.1 \\ 
LMC0219 & $5\uph12\upm18\zdot\ups11$ & $-69\arcd17\arcm02\zdot\arcs9$&   19  & 8.0 & 0.05 \\ 
LMC0220 & $5\uph12\upm21\zdot\ups16$ & $-69\arcd24\arcm41\zdot\arcs3$&   14  & 8.05 & 0.05 \\ 
LMC0221 & $5\uph12\upm25\zdot\ups01$ & $-68\arcd46\arcm19\zdot\arcs0$&   34  & 8.6 & 0.05 \\ 
LMC0222 & $5\uph12\upm27\zdot\ups60$ & $-69\arcd33\arcm21\zdot\arcs8$&   11  & 8.9 & 0.05 \\ 
LMC0223 & $5\uph12\upm27\zdot\ups72$ & $-69\arcd21\arcm04\zdot\arcs7$&   31  & 8.75 & 0.05 \\ 
LMC0224 & $5\uph12\upm30\zdot\ups25$ & $-67\arcd17\arcm27\zdot\arcs9$&   16  & 7.0 & 0.1 \\ 
LMC0225 & $5\uph12\upm32\zdot\ups72$ & $-69\arcd13\arcm45\zdot\arcs8$&   16  & 7.8 & 0.1 \\ 
LMC0226 & $5\uph12\upm34\zdot\ups43$ & $-69\arcd17\arcm13\zdot\arcs7$&   16  & 7.6 & 0.08 \\ 
LMC0227 & $5\uph12\upm38\zdot\ups05$ & $-69\arcd17\arcm33\zdot\arcs0$&   14  & 7.9 & 0.1 \\ 
LMC0228 & $5\uph12\upm39\zdot\ups67$ & $-69\arcd10\arcm48\zdot\arcs6$&   16  & 7.8 & 0.05 \\ 
LMC0229 & $5\uph12\upm40\zdot\ups28$ & $-67\arcd37\arcm24\zdot\arcs1$&   25  & 7.9 & 0.05 \\ 
LMC0230 & $5\uph12\upm48\zdot\ups98$ & $-68\arcd51\arcm51\zdot\arcs4$&   18  & 8.0 & 0.05 \\ 
LMC0231 & $5\uph12\upm57\zdot\ups20$ & $-68\arcd56\arcm33\zdot\arcs0$&   23  & 8.0 & 0.08 \\ 
\hline}

\setcounter{table}{0}
\MakeTableSep{|c|c|c|c|c|c|}{12.5cm}{continued}%5
{\hline
Name  &  $\alpha_{2000}$ & $\delta_{2000}$ & Radius
& $\log t$ & $\sigma_{\log t}$\\
OGLE-CL-& & & [\arcs] & & \\
\hline\xrule
LMC0232 & $5\uph12\upm57\zdot\ups60$ & $-69\arcd04\arcm05\zdot\arcs7$&   12  & 8.25 & 0.05 \\ 
LMC0233 & $5\uph13\upm03\zdot\ups60$ & $-69\arcd02\arcm59\zdot\arcs6$&   16  & 9.05 & 0.05 \\ 
LMC0234 & $5\uph13\upm07\zdot\ups87$ & $-69\arcd26\arcm58\zdot\arcs0$&   26  & 7.8 & 0.1 \\ 
LMC0235 & $5\uph13\upm10\zdot\ups88$ & $-69\arcd07\arcm02\zdot\arcs9$&   28  & 7.95 & 0.05 \\ 
LMC0236 & $5\uph13\upm11\zdot\ups65$ & $-69\arcd18\arcm45\zdot\arcs0$&   43  & 8.5 & 0.1 \\ 
LMC0237 & $5\uph13\upm13\zdot\ups22$ & $-69\arcd22\arcm30\zdot\arcs3$&   18  & $<$6.7 & -- \\ 
LMC0238 & $5\uph13\upm19\zdot\ups04$ & $-69\arcd21\arcm44\zdot\arcs5$&   32  & $<$6.7 & -- \\ 
LMC0239 & $5\uph13\upm19\zdot\ups34$ & $-69\arcd12\arcm33\zdot\arcs8$&   14  & 8.55 & 0.05 \\ 
LMC0240 & $5\uph13\upm21\zdot\ups75$ & $-69\arcd22\arcm37\zdot\arcs9$&   27  & 7.0 & 0.1 \\ 
LMC0241 & $5\uph13\upm25\zdot\ups65$ & $-69\arcd10\arcm50\zdot\arcs1$&   18  & 6.7 & 0.05 \\ 
LMC0242 & $5\uph13\upm28\zdot\ups42$ & $-69\arcd22\arcm21\zdot\arcs7$&   23  & $<$6.7 & -- \\ 
LMC0243 & $5\uph13\upm29\zdot\ups40$ & $-69\arcd27\arcm57\zdot\arcs2$&   23  & 8.8 & 0.1 \\ 
LMC0244 & $5\uph13\upm35\zdot\ups75$ & $-68\arcd49\arcm28\zdot\arcs3$&   18  & 7.6 & 0.1 \\ 
LMC0245 & $5\uph13\upm37\zdot\ups33$ & $-69\arcd18\arcm03\zdot\arcs3$&   23  & $<$6.7 & -- \\ 
LMC0246 & $5\uph13\upm38\zdot\ups90$ & $-69\arcd23\arcm02\zdot\arcs0$&   18  & $<$6.8 & -- \\ 
LMC0247 & $5\uph13\upm40\zdot\ups08$ & $-69\arcd22\arcm26\zdot\arcs8$&   11  & 7.8 & 0.1 \\ 
LMC0248 & $5\uph13\upm50\zdot\ups85$ & $-69\arcd29\arcm37\zdot\arcs9$&   12  & 7.8 & 0.1 \\ 
LMC0249 & $5\uph13\upm53\zdot\ups97$ & $-69\arcd24\arcm18\zdot\arcs9$&   12  & 8.4 & 0.1 \\ 
LMC0251 & $5\uph14\upm01\zdot\ups55$ & $-68\arcd56\arcm59\zdot\arcs1$&   12  & 8.2 & 0.05 \\ 
LMC0253 & $5\uph14\upm17\zdot\ups64$ & $-69\arcd06\arcm00\zdot\arcs9$&   12  & 8.5 & 0.08 \\ 
LMC0254 & $5\uph14\upm39\zdot\ups80$ & $-68\arcd57\arcm48\zdot\arcs2$&   13  & 8.0 & 0.1 \\ 
LMC0255 & $5\uph14\upm48\zdot\ups29$ & $-68\arcd54\arcm21\zdot\arcs6$&   24  & 8.1 & 0.05 \\ 
LMC0257 & $5\uph14\upm51\zdot\ups14$ & $-69\arcd25\arcm46\zdot\arcs8$&   11  & 8.75 & 0.1 \\ 
LMC0259 & $5\uph15\upm02\zdot\ups34$ & $-69\arcd20\arcm08\zdot\arcs3$&   23  & 8.7 & 0.07 \\ 
LMC0261 & $5\uph15\upm06\zdot\ups95$ & $-68\arcd58\arcm43\zdot\arcs4$&   33  & 8.1 & 0.03 \\ 
LMC0262 & $5\uph15\upm14\zdot\ups62$ & $-68\arcd52\arcm57\zdot\arcs2$&   22  & 8.1 & 0.1 \\ 
LMC0265 & $5\uph15\upm25\zdot\ups80$ & $-69\arcd03\arcm02\zdot\arcs7$&   18  & 8.2 & 0.1 \\ 
LMC0266 & $5\uph15\upm27\zdot\ups32$ & $-69\arcd20\arcm43\zdot\arcs0$&   18  & 8.0 & 0.1 \\ 
LMC0268 & $5\uph15\upm34\zdot\ups64$ & $-69\arcd34\arcm42\zdot\arcs0$&    8  & 7.5 & 0.1 \\ 
LMC0269 & $5\uph15\upm35\zdot\ups62$ & $-69\arcd08\arcm20\zdot\arcs8$&   29  & 8.05 & 0.8 \\ 
LMC0270 & $5\uph15\upm37\zdot\ups18$ & $-69\arcd28\arcm24\zdot\arcs5$&   25  & 6.7 & 0.1 \\ 
LMC0272 & $5\uph15\upm39\zdot\ups48$ & $-69\arcd37\arcm40\zdot\arcs1$&   14  & 8.9 & 0.05 \\ 
LMC0273 & $5\uph15\upm40\zdot\ups26$ & $-69\arcd16\arcm50\zdot\arcs7$&   45  & 7.55 & 0.05 \\ 
LMC0274 & $5\uph15\upm40\zdot\ups46$ & $-69\arcd20\arcm18\zdot\arcs2$&   13  & 7.95 & 0.05 \\ 
LMC0276 & $5\uph15\upm46\zdot\ups50$ & $-69\arcd14\arcm39\zdot\arcs2$&   18  & 7.75 & 0.05 \\ 
LMC0278 & $5\uph15\upm52\zdot\ups01$ & $-69\arcd28\arcm08\zdot\arcs2$&   31  & 7.85 & 0.05 \\ 
LMC0281 & $5\uph16\upm01\zdot\ups40$ & $-69\arcd24\arcm47\zdot\arcs3$&   41  & 7.9 & 0.1 \\ 
LMC0282 & $5\uph16\upm03\zdot\ups53$ & $-69\arcd06\arcm09\zdot\arcs2$&   18  & 8.45 & 0.05 \\ 
LMC0284 & $5\uph16\upm16\zdot\ups04$ & $-69\arcd26\arcm19\zdot\arcs9$&   15  & 7.6 & 0.1 \\ 
LMC0285 & $5\uph16\upm16\zdot\ups84$ & $-69\arcd09\arcm15\zdot\arcs2$&   11  & 8.1 & 0.1 \\ 
LMC0286 & $5\uph16\upm21\zdot\ups01$ & $-69\arcd32\arcm31\zdot\arcs1$&   18  & $<$6.8 & -- \\ 
LMC0289 & $5\uph16\upm26\zdot\ups92$ & $-69\arcd40\arcm26\zdot\arcs9$&   11  & 7.8 & 0.1 \\ 
LMC0290 & $5\uph16\upm32\zdot\ups11$ & $-68\arcd55\arcm07\zdot\arcs4$&   17  & 7.7 & 0.1 \\ 
LMC0292 & $5\uph16\upm41\zdot\ups24$ & $-69\arcd39\arcm24\zdot\arcs4$&   35  & $>$9.2 & -- \\ 
LMC0293 & $5\uph16\upm44\zdot\ups63$ & $-69\arcd27\arcm42\zdot\arcs9$&   12  & 7.3 & 0.2 \\ 
LMC0295 & $5\uph16\upm49\zdot\ups56$ & $-69\arcd29\arcm50\zdot\arcs6$&   12  & $<$6.7 & -- \\ 
LMC0298 & $5\uph16\upm52\zdot\ups88$ & $-69\arcd09\arcm00\zdot\arcs0$&   12  & 7.3 & 0.2 \\ 
LMC0299 & $5\uph16\upm53\zdot\ups02$ & $-69\arcd25\arcm11\zdot\arcs5$&   20  & 8.1 & 0.05 \\ 
LMC0300 & $5\uph16\upm53\zdot\ups42$ & $-69\arcd43\arcm27\zdot\arcs0$&   11  & 9.05 & 0.1 \\ 
LMC0301 & $5\uph16\upm54\zdot\ups05$ & $-69\arcd34\arcm56\zdot\arcs3$&   14  & 8.2 & 0.1 \\ 
\hline}

\clearpage
\setcounter{table}{0}
\MakeTableSep{|c|c|c|c|c|c|}{12.5cm}{continued}%6
{\hline
Name  &  $\alpha_{2000}$ & $\delta_{2000}$ & Radius
& $\log t$ & $\sigma_{\log t}$\\
OGLE-CL-& & & [\arcs] & & \\
\hline\xrule
LMC0302 & $5\uph16\upm54\zdot\ups41$ & $-68\arcd52\arcm35\zdot\arcs8$&   20  & 8.65 & 0.05 \\ 
LMC0303 & $5\uph16\upm55\zdot\ups59$ & $-69\arcd08\arcm51\zdot\arcs2$&   30  & 6.7 & 0.1 \\ 
LMC0304 & $5\uph17\upm08\zdot\ups00$ & $-68\arcd52\arcm23\zdot\arcs5$&   41  & 9.0 & 0.1 \\ 
LMC0305 & $5\uph17\upm14\zdot\ups76$ & $-69\arcd32\arcm26\zdot\arcs3$&   39  & 6.8 & 0.15 \\ 
LMC0307 & $5\uph17\upm19\zdot\ups96$ & $-69\arcd12\arcm48\zdot\arcs5$&   18  & 9.1 & 0.08 \\ 
LMC0309 & $5\uph17\upm22\zdot\ups39$ & $-69\arcd20\arcm16\zdot\arcs2$&   49  & 7.88 & 0.03 \\ 
LMC0310 & $5\uph17\upm25\zdot\ups62$ & $-69\arcd06\arcm54\zdot\arcs5$&   39  & $<$6.7 & -- \\ 
LMC0311 & $5\uph17\upm26\zdot\ups59$ & $-69\arcd22\arcm31\zdot\arcs8$&   45  & 9.0 & 0.08 \\ 
LMC0312 & $5\uph17\upm27\zdot\ups68$ & $-69\arcd21\arcm22\zdot\arcs3$&   31  & 8.0 & 0.05 \\ 
LMC0313 & $5\uph17\upm29\zdot\ups24$ & $-69\arcd24\arcm58\zdot\arcs0$&   11  & 8.25 & 0.05 \\ 
LMC0314 & $5\uph17\upm33\zdot\ups42$ & $-69\arcd30\arcm53\zdot\arcs2$&   24  & 8.5 & 0.05 \\ 
LMC0315 & $5\uph17\upm38\zdot\ups26$ & $-68\arcd58\arcm21\zdot\arcs5$&   24  & 8.5 & 0.1 \\ 
LMC0316 & $5\uph17\upm43\zdot\ups83$ & $-69\arcd34\arcm06\zdot\arcs1$&   22  & 6.8 & 0.2 \\ 
LMC0318 & $5\uph17\upm48\zdot\ups65$ & $-69\arcd38\arcm40\zdot\arcs5$&   43  & 9.05 & 0.1 \\ 
LMC0319 & $5\uph17\upm48\zdot\ups72$ & $-69\arcd24\arcm36\zdot\arcs2$&   26  & 7.0 & 0.1 \\ 
LMC0320 & $5\uph17\upm49\zdot\ups82$ & $-69\arcd41\arcm39\zdot\arcs5$&   10  & 7.65 & 0.05 \\ 
LMC0321 & $5\uph17\upm56\zdot\ups16$ & $-69\arcd34\arcm52\zdot\arcs3$&   17  & 8.1 & 0.07 \\ 
LMC0322 & $5\uph18\upm00\zdot\ups61$ & $-69\arcd08\arcm00\zdot\arcs6$&   12  & 9.0 & 0.05 \\ 
LMC0323 & $5\uph18\upm05\zdot\ups10$ & $-69\arcd10\arcm17\zdot\arcs8$&   24  & 7.9 & 0.1 \\ 
LMC0324 & $5\uph18\upm06\zdot\ups44$ & $-69\arcd31\arcm46\zdot\arcs4$&   29  & 7.4 & 0.1 \\ 
LMC0326 & $5\uph18\upm10\zdot\ups51$ & $-69\arcd32\arcm26\zdot\arcs8$&   14  & 8.0 & 0.1 \\ 
LMC0327 & $5\uph18\upm10\zdot\ups88$ & $-69\arcd16\arcm52\zdot\arcs7$&   23  & $<$6.6 & -- \\ 
LMC0328 & $5\uph18\upm11\zdot\ups41$ & $-69\arcd13\arcm05\zdot\arcs7$&   32  & $<$6.6 & -- \\ 
LMC0329 & $5\uph18\upm18\zdot\ups05$ & $-69\arcd45\arcm04\zdot\arcs9$&   18  & 8.0 & 0.1 \\ 
LMC0330 & $5\uph18\upm18\zdot\ups74$ & $-69\arcd32\arcm14\zdot\arcs8$&   41  & 7.38 & 0.03 \\ 
LMC0331 & $5\uph18\upm24\zdot\ups72$ & $-69\arcd29\arcm05\zdot\arcs8$&   33  & 7.4 & 0.05 \\ 
LMC0332 & $5\uph18\upm25\zdot\ups56$ & $-69\arcd19\arcm30\zdot\arcs2$&   32  & 8.1,8.4 & 0.05 \\ 
LMC0333 & $5\uph18\upm28\zdot\ups89$ & $-69\arcd37\arcm00\zdot\arcs1$&   39  & 7.0 & 0.05 \\ 
LMC0334 & $5\uph18\upm31\zdot\ups19$ & $-69\arcd45\arcm14\zdot\arcs6$&   24  & 7.95 & 0.08 \\ 
LMC0336 & $5\uph18\upm37\zdot\ups87$ & $-69\arcd24\arcm22\zdot\arcs9$&   61  & $>$9.2 & -- \\ 
LMC0337 & $5\uph18\upm41\zdot\ups31$ & $-69\arcd04\arcm46\zdot\arcs1$&   18  & 8.8 & 0.1 \\ 
LMC0338 & $5\uph18\upm42\zdot\ups53$ & $-69\arcd14\arcm12\zdot\arcs3$&   46  & 6.7 & 0.1 \\ 
LMC0340 & $5\uph18\upm46\zdot\ups72$ & $-69\arcd13\arcm32\zdot\arcs4$&   19  & $<$6.8 & -- \\ 
LMC0341 & $5\uph18\upm51\zdot\ups25$ & $-69\arcd22\arcm14\zdot\arcs1$&   30  & 8.9 & 0.05 \\ 
LMC0342 & $5\uph18\upm53\zdot\ups28$ & $-69\arcd31\arcm19\zdot\arcs3$&   24  & 8.0 & 0.05 \\ 
LMC0343 & $5\uph19\upm02\zdot\ups10$ & $-69\arcd00\arcm03\zdot\arcs8$&   37  & 9.0 & 0.05 \\ 
LMC0344 & $5\uph19\upm03\zdot\ups76$ & $-69\arcd11\arcm36\zdot\arcs1$&   30  & $<$6.7 & -- \\ 
LMC0345 & $5\uph19\upm04\zdot\ups55$ & $-69\arcd48\arcm38\zdot\arcs3$&   18  & 8.8 & 0.03 \\ 
LMC0346 & $5\uph19\upm08\zdot\ups90$ & $-69\arcd15\arcm36\zdot\arcs2$&   13  & 8.75 & 0.05 \\ 
LMC0347 & $5\uph19\upm18\zdot\ups99$ & $-69\arcd42\arcm51\zdot\arcs1$&   14  & 8.3 & 0.08 \\ 
LMC0348 & $5\uph19\upm24\zdot\ups00$ & $-69\arcd47\arcm16\zdot\arcs5$&   41  & 8.1 & 0.1 \\ 
LMC0349 & $5\uph19\upm24\zdot\ups09$ & $-69\arcd39\arcm01\zdot\arcs1$&   41  & 6.6 & 0.1 \\ 
LMC0350 & $5\uph19\upm24\zdot\ups78$ & $-69\arcd19\arcm18\zdot\arcs0$&   36  & 8.5 & 0.1 \\ 
LMC0351 & $5\uph19\upm25\zdot\ups74$ & $-69\arcd32\arcm27\zdot\arcs1$&   33  & 8.3 & 0.05 \\ 
LMC0352 & $5\uph19\upm33\zdot\ups20$ & $-69\arcd26\arcm44\zdot\arcs5$&   18  & 8.7 & 0.05 \\ 
LMC0353 & $5\uph19\upm33\zdot\ups88$ & $-69\arcd32\arcm31\zdot\arcs9$&   26  & 8.8 & 0.1 \\ 
LMC0354 & $5\uph19\upm49\zdot\ups29$ & $-69\arcd29\arcm41\zdot\arcs7$&   31  & 7.15 & 0.1 \\ 
LMC0355 & $5\uph19\upm49\zdot\ups52$ & $-69\arcd26\arcm56\zdot\arcs5$&   30  & $<$6.7 & -- \\ 
LMC0356 & $5\uph19\upm54\zdot\ups37$ & $-68\arcd57\arcm52\zdot\arcs7$&   30  & 9.0 & 0.03 \\ 
LMC0358 & $5\uph19\upm57\zdot\ups35$ & $-69\arcd41\arcm27\zdot\arcs6$&   16  & 8.0 & 0.1 \\ 
\hline}

\setcounter{table}{0}
\MakeTableSep{|c|c|c|c|c|c|}{12.5cm}{continued}%7
{\hline
Name  &  $\alpha_{2000}$ & $\delta_{2000}$ & Radius
& $\log t$ & $\sigma_{\log t}$\\
OGLE-CL-& & & [\arcs] & & \\
\hline\xrule
LMC0359 & $5\uph19\upm57\zdot\ups48$ & $-69\arcd25\arcm02\zdot\arcs8$&   18  & 8.0 & 0.1 \\ 
LMC0360 & $5\uph19\upm59\zdot\ups59$ & $-70\arcd39\arcm54\zdot\arcs6$&   17  & 8.4 & 0.1 \\ 
LMC0361 & $5\uph20\upm02\zdot\ups05$ & $-69\arcd15\arcm39\zdot\arcs6$&   10  & 8.0 & 0.1 \\ 
LMC0362 & $5\uph20\upm03\zdot\ups01$ & $-69\arcd23\arcm59\zdot\arcs1$&   11  & 9 & 0.05 \\ 
LMC0365 & $5\uph20\upm08\zdot\ups04$ & $-70\arcd09\arcm15\zdot\arcs0$&   10  & 8.3 & 0.1 \\ 
LMC0367 & $5\uph20\upm15\zdot\ups93$ & $-69\arcd20\arcm24\zdot\arcs8$&   14  & 8.0 & 0.1 \\ 
LMC0368 & $5\uph20\upm20\zdot\ups91$ & $-70\arcd46\arcm06\zdot\arcs2$&   18  & 8.1 & 0.05 \\ 
LMC0369 & $5\uph20\upm23\zdot\ups57$ & $-69\arcd35\arcm03\zdot\arcs1$&   28  & 8.3 & 0.07 \\ 
LMC0370 & $5\uph20\upm25\zdot\ups45$ & $-69\arcd21\arcm18\zdot\arcs1$&   20  & 7.8 & 0.1 \\ 
LMC0372 & $5\uph20\upm27\zdot\ups62$ & $-69\arcd21\arcm53\zdot\arcs3$&   20  & 8.55 & 0.1 \\ 
LMC0373 & $5\uph20\upm27\zdot\ups71$ & $-70\arcd27\arcm00\zdot\arcs0$&   15  & 8.55 & 0.05 \\ 
LMC0374 & $5\uph20\upm29\zdot\ups05$ & $-69\arcd44\arcm59\zdot\arcs5$&   10  & 7.8 & 0.1 \\ 
LMC0375 & $5\uph20\upm30\zdot\ups61$ & $-69\arcd32\arcm09\zdot\arcs0$&   31  & 7.7 & 0.05 \\ 
LMC0376 & $5\uph20\upm34\zdot\ups23$ & $-69\arcd38\arcm18\zdot\arcs2$&   22  & 6.8 & 0.1 \\ 
LMC0377 & $5\uph20\upm34\zdot\ups71$ & $-70\arcd00\arcm53\zdot\arcs0$&   16  & 7.55 & 0.05 \\ 
LMC0378 & $5\uph20\upm35\zdot\ups06$ & $-69\arcd41\arcm19\zdot\arcs4$&   18  & 8.7 & 0.1 \\ 
LMC0379 & $5\uph20\upm35\zdot\ups42$ & $-69\arcd31\arcm32\zdot\arcs9$&   27  & 8.0 & 0.1 \\ 
LMC0380 & $5\uph20\upm37\zdot\ups00$ & $-70\arcd57\arcm51\zdot\arcs7$&    9  & 9 & 0.1 \\ 
LMC0381 & $5\uph20\upm48\zdot\ups15$ & $-69\arcd24\arcm55\zdot\arcs2$&   16  & 6.8 & 0.1 \\ 
LMC0382 & $5\uph20\upm57\zdot\ups73$ & $-69\arcd28\arcm40\zdot\arcs2$&   31  & 9.03 & 0.05 \\ 
LMC0384 & $5\uph21\upm00\zdot\ups04$ & $-70\arcd18\arcm57\zdot\arcs4$&   12  & 8.8 & 0.1 \\ 
LMC0385 & $5\uph21\upm01\zdot\ups62$ & $-69\arcd23\arcm22\zdot\arcs6$&   20  & 8.0 & 0.05 \\ 
LMC0386 & $5\uph21\upm02\zdot\ups07$ & $-70\arcd52\arcm24\zdot\arcs1$&   15  & 8.3 & 0.05 \\ 
LMC0387 & $5\uph21\upm05\zdot\ups47$ & $-70\arcd02\arcm45\zdot\arcs3$&   12  & 8.0 & 0.08 \\ 
LMC0388 & $5\uph21\upm09\zdot\ups93$ & $-69\arcd50\arcm36\zdot\arcs4$&   12  & 9.0 & 0.1 \\ 
LMC0389 & $5\uph21\upm10\zdot\ups93$ & $-69\arcd56\arcm36\zdot\arcs8$&   16  & 8.7 & 0.08 \\ 
LMC0390 & $5\uph21\upm18\zdot\ups65$ & $-69\arcd28\arcm35\zdot\arcs7$&   17  & 9.0 & 0.08 \\ 
LMC0391 & $5\uph21\upm21\zdot\ups69$ & $-70\arcd54\arcm01\zdot\arcs0$&   17  & 9.0 & 0.1 \\ 
LMC0392 & $5\uph21\upm22\zdot\ups77$ & $-69\arcd54\arcm33\zdot\arcs5$&   12  & 8.6 & 0.08 \\ 
LMC0394 & $5\uph21\upm24\zdot\ups45$ & $-69\arcd56\arcm27\zdot\arcs5$&   25  & 8.55 & 0.05 \\ 
LMC0395 & $5\uph21\upm26\zdot\ups82$ & $-69\arcd56\arcm59\zdot\arcs0$&   30  & 9.0 & 0.1 \\ 
LMC0396 & $5\uph21\upm29\zdot\ups91$ & $-69\arcd49\arcm43\zdot\arcs0$&   30  & 8.0 & 0.1 \\ 
LMC0397 & $5\uph21\upm30\zdot\ups04$ & $-69\arcd25\arcm54\zdot\arcs6$&   11  & 6.6,8.1 & 0.1 \\ 
LMC0398 & $5\uph21\upm35\zdot\ups12$ & $-69\arcd40\arcm20\zdot\arcs8$&   18  & 7.45 & 0.05 \\ 
LMC0399 & $5\uph21\upm47\zdot\ups36$ & $-69\arcd24\arcm59\zdot\arcs8$&   14  & 8.5 & 0.1 \\ 
LMC0400 & $5\uph21\upm49\zdot\ups43$ & $-69\arcd39\arcm06\zdot\arcs0$&   30  & 8.2 & 0.2 \\ 
LMC0402 & $5\uph21\upm57\zdot\ups00$ & $-69\arcd36\arcm39\zdot\arcs0$&   12  & 8.4 & 0.1 \\ 
LMC0403 & $5\uph22\upm03\zdot\ups23$ & $-70\arcd02\arcm44\zdot\arcs3$&   12  & 8.25 & 0.05 \\ 
LMC0404 & $5\uph22\upm03\zdot\ups30$ & $-69\arcd15\arcm17\zdot\arcs9$&   24  & 8.35 & 0.05 \\ 
LMC0405 & $5\uph22\upm06\zdot\ups85$ & $-69\arcd14\arcm44\zdot\arcs7$&   14  & 8.3 & 0.1 \\ 
LMC0407 & $5\uph22\upm14\zdot\ups67$ & $-69\arcd30\arcm40\zdot\arcs7$&   40  & 8.2 & 0.1 \\ 
LMC0409 & $5\uph22\upm27\zdot\ups28$ & $-69\arcd44\arcm43\zdot\arcs0$&   24  & 7.55 & 0.05 \\ 
LMC0411 & $5\uph22\upm29\zdot\ups67$ & $-70\arcd09\arcm17\zdot\arcs0$&   30  & 8.15 & 0.05 \\ 
LMC0412 & $5\uph22\upm32\zdot\ups79$ & $-69\arcd33\arcm02\zdot\arcs0$&   23  & 8.1,8.6 & 0.1 \\ 
LMC0413 & $5\uph22\upm37\zdot\ups90$ & $-69\arcd44\arcm39\zdot\arcs9$&   22  & 8.75 & 0.05 \\ 
LMC0414 & $5\uph22\upm45\zdot\ups25$ & $-69\arcd54\arcm06\zdot\arcs6$&   14  & 8.0 & 0.05 \\ 
LMC0415 & $5\uph22\upm57\zdot\ups78$ & $-69\arcd37\arcm30\zdot\arcs7$&   14  & 7.5 & 0.1 \\ 
LMC0416 & $5\uph23\upm10\zdot\ups33$ & $-69\arcd52\arcm01\zdot\arcs4$&   20  & 7.75 & 0.05 \\ 
LMC0417 & $5\uph23\upm12\zdot\ups94$ & $-69\arcd49\arcm23\zdot\arcs0$&   22  & 8.3 & 0.1 \\ 
LMC0419 & $5\uph23\upm25\zdot\ups24$ & $-69\arcd50\arcm07\zdot\arcs1$&   26  & 7.98 & 0.05 \\ 
\hline}

\setcounter{table}{0}
\MakeTableSep{|c|c|c|c|c|c|}{12.5cm}{continued}%8
{\hline
Name  &  $\alpha_{2000}$ & $\delta_{2000}$ & Radius
& $\log t$ & $\sigma_{\log t}$\\
OGLE-CL-& & & [\arcs] & & \\
\hline\xrule
LMC0420 & $5\uph23\upm27\zdot\ups48$ & $-69\arcd45\arcm06\zdot\arcs4$&   17  & 8.9 & 0.05 \\ 
LMC0422 & $5\uph23\upm32\zdot\ups17$ & $-69\arcd54\arcm14\zdot\arcs0$&   11  & 8.2 & 0.1 \\ 
LMC0423 & $5\uph23\upm32\zdot\ups83$ & $-69\arcd20\arcm33\zdot\arcs9$&   29  & 9.0 & 0.08 \\ 
LMC0424 & $5\uph23\upm35\zdot\ups48$ & $-69\arcd54\arcm17\zdot\arcs7$&   10  & 8.0 & 0.1 \\ 
LMC0426 & $5\uph23\upm39\zdot\ups33$ & $-69\arcd14\arcm42\zdot\arcs4$&   34  & 9.1 & 0.1 \\ 
LMC0428 & $5\uph23\upm58\zdot\ups40$ & $-69\arcd57\arcm25\zdot\arcs5$&   10  & 9.0 & 0.1 \\ 
LMC0429 & $5\uph24\upm06\zdot\ups96$ & $-69\arcd44\arcm26\zdot\arcs7$&   14  & 8.3 & 0.05 \\ 
LMC0430 & $5\uph24\upm16\zdot\ups37$ & $-69\arcd39\arcm12\zdot\arcs9$&   12  & 8.8 & 0.05 \\ 
LMC0431 & $5\uph24\upm20\zdot\ups42$ & $-69\arcd46\arcm26\zdot\arcs4$&   19  & 8.0 & 0.05 \\ 
LMC0432 & $5\uph24\upm21\zdot\ups11$ & $-69\arcd57\arcm55\zdot\arcs0$&   24  & 8.4 & 0.1 \\ 
LMC0433 & $5\uph24\upm21\zdot\ups58$ & $-69\arcd38\arcm28\zdot\arcs9$&   11  & 9.0 & 0.1 \\ 
LMC0434 & $5\uph24\upm23\zdot\ups94$ & $-69\arcd46\arcm47\zdot\arcs5$&   10  & 9.0 & 0.1 \\ 
LMC0436 & $5\uph24\upm33\zdot\ups04$ & $-69\arcd54\arcm04\zdot\arcs3$&   44  & 8.7 & 0.08 \\ 
LMC0437 & $5\uph24\upm33\zdot\ups45$ & $-69\arcd55\arcm26\zdot\arcs9$&   14  & 8.7 & 0.1 \\ 
LMC0438 & $5\uph24\upm33\zdot\ups50$ & $-69\arcd44\arcm43\zdot\arcs1$&   36  & 8.6 & 0.08 \\ 
LMC0439 & $5\uph24\upm41\zdot\ups16$ & $-69\arcd41\arcm34\zdot\arcs2$&   11  & 8.7 & 0.1 \\ 
LMC0440 & $5\uph24\upm41\zdot\ups59$ & $-69\arcd53\arcm10\zdot\arcs8$&   21  & 8.1 & 0.08 \\ 
LMC0442 & $5\uph24\upm53\zdot\ups02$ & $-69\arcd49\arcm47\zdot\arcs2$&   29  & 8.1 & 0.1 \\ 
LMC0443 & $5\uph24\upm55\zdot\ups33$ & $-69\arcd50\arcm13\zdot\arcs9$&   22  & 8.0 & 0.03 \\ 
LMC0444 & $5\uph24\upm55\zdot\ups46$ & $-69\arcd51\arcm46\zdot\arcs0$&   14  & 8.0 & 0.08 \\ 
LMC0445 & $5\uph24\upm56\zdot\ups68$ & $-69\arcd25\arcm29\zdot\arcs3$&   18  & 8.6 & 0.1 \\ 
LMC0446 & $5\uph25\upm01\zdot\ups13$ & $-69\arcd26\arcm03\zdot\arcs1$&   35  & 8.3 & 0.05 \\ 
LMC0447 & $5\uph25\upm03\zdot\ups54$ & $-69\arcd52\arcm12\zdot\arcs7$&   12  & 8.1 & 0.1 \\ 
LMC0448 & $5\uph25\upm04\zdot\ups69$ & $-69\arcd44\arcm14\zdot\arcs3$&   20  & 8.6 & 0.1 \\ 
LMC0450 & $5\uph25\upm06\zdot\ups87$ & $-69\arcd42\arcm56\zdot\arcs3$&   20  & 8.2 & 0.1 \\ 
LMC0451 & $5\uph25\upm14\zdot\ups78$ & $-70\arcd05\arcm57\zdot\arcs3$&   12  & 8.1 & 0.1 \\ 
LMC0452 & $5\uph25\upm17\zdot\ups64$ & $-69\arcd32\arcm23\zdot\arcs0$&   14  & 8.9 & 0.05 \\ 
LMC0453 & $5\uph25\upm22\zdot\ups80$ & $-69\arcd26\arcm23\zdot\arcs3$&   12  & $<$7 & -- \\ 
LMC0454 & $5\uph25\upm23\zdot\ups00$ & $-69\arcd47\arcm07\zdot\arcs0$&   16  & 8.6,9.0 & 0.05 \\ 
LMC0456 & $5\uph25\upm28\zdot\ups00$ & $-69\arcd46\arcm31\zdot\arcs6$&   21  & 8.45 & 0.05 \\ 
LMC0457 & $5\uph25\upm30\zdot\ups72$ & $-69\arcd50\arcm09\zdot\arcs6$&   37  & 8.0 & 0.1 \\ 
LMC0459 & $5\uph25\upm35\zdot\ups72$ & $-69\arcd55\arcm35\zdot\arcs6$&   33  & 8.7 & 0.1 \\ 
LMC0460 & $5\uph25\upm38\zdot\ups18$ & $-70\arcd15\arcm42\zdot\arcs2$&   14  & 8.2 & 0.1 \\ 
LMC0461 & $5\uph25\upm38\zdot\ups49$ & $-69\arcd49\arcm30\zdot\arcs8$&   30  & 7.95 & 0.05 \\ 
LMC0463 & $5\uph25\upm49\zdot\ups91$ & $-69\arcd38\arcm28\zdot\arcs2$&   14  & 8.5 & 0.1 \\ 
LMC0465 & $5\uph25\upm53\zdot\ups77$ & $-69\arcd46\arcm13\zdot\arcs5$&   20  & 8.8 & 0.05 \\ 
LMC0467 & $5\uph25\upm57\zdot\ups30$ & $-69\arcd45\arcm03\zdot\arcs9$&   22  & 8.3 & 0.05 \\ 
LMC0468 & $5\uph26\upm01\zdot\ups94$ & $-69\arcd30\arcm20\zdot\arcs6$&   29  & 8.3 & 0.05 \\ 
LMC0469 & $5\uph26\upm05\zdot\ups49$ & $-70\arcd05\arcm34\zdot\arcs3$&   19  & 8.6 & 0.1 \\ 
LMC0470 & $5\uph26\upm14\zdot\ups40$ & $-69\arcd33\arcm57\zdot\arcs1$&   16  & 8.9 & 0.05 \\ 
LMC0471 & $5\uph26\upm17\zdot\ups43$ & $-70\arcd13\arcm16\zdot\arcs4$&   24  & 8.85 & 0.05 \\ 
LMC0472 & $5\uph26\upm19\zdot\ups41$ & $-69\arcd30\arcm06\zdot\arcs3$&   18  & 6.7 & 0.1 \\ 
LMC0473 & $5\uph26\upm23\zdot\ups96$ & $-69\arcd43\arcm50\zdot\arcs9$&   24  & 8.4 & 0.1 \\ 
LMC0474 & $5\uph26\upm24\zdot\ups98$ & $-69\arcd40\arcm57\zdot\arcs8$&   17  & 8.9 & 0.05 \\ 
LMC0476 & $5\uph26\upm33\zdot\ups08$ & $-69\arcd48\arcm12\zdot\arcs0$&   23  & 8.4 & 0.1 \\ 
LMC0477 & $5\uph26\upm34\zdot\ups11$ & $-69\arcd50\arcm26\zdot\arcs7$&   37  & 7.8 & 0.05 \\ 
LMC0478 & $5\uph26\upm35\zdot\ups30$ & $-69\arcd49\arcm23\zdot\arcs1$&   30  & 8.0 & 0.1 \\ 
LMC0479 & $5\uph26\upm35\zdot\ups88$ & $-69\arcd36\arcm52\zdot\arcs6$&   23  & 7.75,8.5 & 0.08 \\ 
LMC0480 & $5\uph26\upm45\zdot\ups58$ & $-69\arcd51\arcm03\zdot\arcs2$&   31  & 8.0 & 0.05 \\ 
LMC0481 & $5\uph26\upm48\zdot\ups80$ & $-69\arcd50\arcm17\zdot\arcs2$&   29  & 7.8 & 0.1 \\ 
\hline}

\setcounter{table}{0}
\MakeTableSep{|c|c|c|c|c|c|}{12.5cm}{continued}%9
{\hline
Name  &  $\alpha_{2000}$ & $\delta_{2000}$ & Radius
& $\log t$ & $\sigma_{\log t}$\\
OGLE-CL-& & & [\arcs] & & \\
\hline\xrule
LMC0482 & $5\uph26\upm52\zdot\ups66$ & $-69\arcd46\arcm03\zdot\arcs0$&   22  & 8.9 & 9.0 \\ 
LMC0483 & $5\uph26\upm53\zdot\ups23$ & $-69\arcd48\arcm53\zdot\arcs5$&   18  & 8.8 & 0.1 \\ 
LMC0484 & $5\uph26\upm53\zdot\ups23$ & $-70\arcd12\arcm33\zdot\arcs2$&   38  & 8.2 & 0.05 \\ 
LMC0485 & $5\uph27\upm00\zdot\ups68$ & $-69\arcd46\arcm37\zdot\arcs5$&   25  & 7.95 & 0.05 \\ 
LMC0486 & $5\uph27\upm00\zdot\ups70$ & $-69\arcd42\arcm37\zdot\arcs9$&   25  & 8.7 & 0.06 \\ 
LMC0487 & $5\uph27\upm04\zdot\ups27$ & $-69\arcd51\arcm51\zdot\arcs5$&   37  & 8.0 & 0.1 \\ 
LMC0488 & $5\uph27\upm07\zdot\ups14$ & $-69\arcd30\arcm55\zdot\arcs2$&   18  & 8.9 & 0.1 \\ 
LMC0490 & $5\uph27\upm21\zdot\ups23$ & $-70\arcd00\arcm41\zdot\arcs3$&   20  & 8.6 & 0.1 \\ 
LMC0491 & $5\uph27\upm22\zdot\ups39$ & $-69\arcd52\arcm12\zdot\arcs7$&   25  & 8.0 & 0.05 \\ 
LMC0493 & $5\uph27\upm29\zdot\ups70$ & $-69\arcd29\arcm24\zdot\arcs3$&   13  & 9.2 & 0.1 \\ 
LMC0495 & $5\uph27\upm35\zdot\ups63$ & $-69\arcd53\arcm49\zdot\arcs6$&   19  & 8.6 & 0.2 \\ 
LMC0496 & $5\uph27\upm37\zdot\ups68$ & $-69\arcd58\arcm13\zdot\arcs9$&   49  & 8.00 & 0.05 \\ 
LMC0497 & $5\uph27\upm47\zdot\ups58$ & $-69\arcd53\arcm29\zdot\arcs8$&   24  & 8.1 & 0.05 \\ 
LMC0498 & $5\uph27\upm59\zdot\ups81$ & $-69\arcd55\arcm50\zdot\arcs4$&   14  & 8.0 & 0.05 \\ 
LMC0499 & $5\uph28\upm03\zdot\ups80$ & $-69\arcd45\arcm54\zdot\arcs6$&   18  & 7.9 & 0.1 \\ 
LMC0500 & $5\uph28\upm05\zdot\ups10$ & $-69\arcd59\arcm16\zdot\arcs7$&   12  & 8.2 & 0.1 \\ 
LMC0502 & $5\uph28\upm10\zdot\ups02$ & $-69\arcd51\arcm29\zdot\arcs2$&   25  & 8.0 & 0.1 \\ 
LMC0503 & $5\uph28\upm16\zdot\ups16$ & $-69\arcd27\arcm23\zdot\arcs0$&   23  & 7.9 & 0.1 \\ 
LMC0504 & $5\uph28\upm25\zdot\ups20$ & $-69\arcd57\arcm12\zdot\arcs0$&   25  & 7.65 & 0.05 \\ 
LMC0505 & $5\uph28\upm26\zdot\ups78$ & $-69\arcd46\arcm05\zdot\arcs3$&   15  & 8.4 & 0.05 \\ 
LMC0506 & $5\uph28\upm27\zdot\ups77$ & $-69\arcd53\arcm48\zdot\arcs5$&   25  & 7.5 & 0.1 \\ 
LMC0507 & $5\uph28\upm31\zdot\ups72$ & $-69\arcd50\arcm32\zdot\arcs1$&   16  & 8.7 & 0.1 \\ 
LMC0508 & $5\uph28\upm35\zdot\ups65$ & $-69\arcd36\arcm39\zdot\arcs0$&   25  & 8.9 & 0.1 \\ 
LMC0510 & $5\uph28\upm41\zdot\ups10$ & $-69\arcd57\arcm13\zdot\arcs0$&   20  & 8.0 & 0.05 \\ 
LMC0511 & $5\uph28\upm42\zdot\ups33$ & $-69\arcd46\arcm06\zdot\arcs4$&   31  & 9.0 & 0.1 \\ 
LMC0512 & $5\uph28\upm44\zdot\ups44$ & $-69\arcd50\arcm04\zdot\arcs9$&   19  & 8.1 & 0.05 \\ 
LMC0513 & $5\uph28\upm51\zdot\ups13$ & $-70\arcd00\arcm41\zdot\arcs8$&   30  & 8.95 & 0.05 \\ 
LMC0515 & $5\uph28\upm54\zdot\ups32$ & $-70\arcd12\arcm18\zdot\arcs2$&   25  & 9.15 & 0.05 \\ 
LMC0516 & $5\uph29\upm05\zdot\ups96$ & $-69\arcd48\arcm30\zdot\arcs0$&   31  & 8.0 & 0.1 \\ 
LMC0517 & $5\uph29\upm18\zdot\ups80$ & $-69\arcd54\arcm52\zdot\arcs5$&   16  & 8.2 & 0.15 \\ 
LMC0518 & $5\uph29\upm19\zdot\ups93$ & $-69\arcd35\arcm56\zdot\arcs5$&   31  & 8.0 & 0.05 \\ 
LMC0519 & $5\uph29\upm23\zdot\ups98$ & $-70\arcd14\arcm12\zdot\arcs0$&   25  & 7.0 & 0.08 \\ 
LMC0520 & $5\uph29\upm24\zdot\ups59$ & $-69\arcd55\arcm11\zdot\arcs8$&   18  & 8.2 & 0.05 \\ 
LMC0523 & $5\uph29\upm32\zdot\ups79$ & $-69\arcd32\arcm33\zdot\arcs6$&   14  & 8.0 & 0.1 \\ 
LMC0524 & $5\uph29\upm33\zdot\ups67$ & $-69\arcd23\arcm22\zdot\arcs0$&   27  & 7.2 & 0.2 \\ 
LMC0525 & $5\uph29\upm34\zdot\ups59$ & $-69\arcd46\arcm32\zdot\arcs8$&   20  & 8.1 & 0.1 \\ 
LMC0526 & $5\uph29\upm34\zdot\ups81$ & $-69\arcd58\arcm31\zdot\arcs6$&   14  & 8.2 & 0.1 \\ 
LMC0528 & $5\uph29\upm53\zdot\ups51$ & $-69\arcd53\arcm23\zdot\arcs0$&   17  & 8.3 & 0.1 \\ 
LMC0530 & $5\uph29\upm59\zdot\ups95$ & $-69\arcd31\arcm21\zdot\arcs3$&   12  & 9.0 & 0.1 \\ 
LMC0531 & $5\uph30\upm00\zdot\ups73$ & $-69\arcd31\arcm37\zdot\arcs1$&   14  & 9.1 & 0.05 \\ 
LMC0532 & $5\uph30\upm01\zdot\ups73$ & $-69\arcd57\arcm02\zdot\arcs3$&   16  & 8.3 & 0.05 \\ 
LMC0534 & $5\uph30\upm03\zdot\ups66$ & $-70\arcd07\arcm32\zdot\arcs2$&    8  & 8.9 & 0.1 \\ 
LMC0535 & $5\uph30\upm03\zdot\ups75$ & $-69\arcd51\arcm19\zdot\arcs8$&   11  & 9.2 & 0.05 \\ 
LMC0536 & $5\uph30\upm03\zdot\ups77$ & $-70\arcd12\arcm15\zdot\arcs4$&   11  & 8.1 & 0.1 \\ 
LMC0538 & $5\uph30\upm10\zdot\ups37$ & $-69\arcd45\arcm09\zdot\arcs6$&   57  & $>$9.2 & -- \\ 
LMC0540 & $5\uph30\upm12\zdot\ups65$ & $-69\arcd47\arcm23\zdot\arcs2$&   45  & 7.9 & 0.1 \\ 
LMC0541 & $5\uph30\upm21\zdot\ups20$ & $-69\arcd35\arcm02\zdot\arcs6$&   31  & 8.2 & 0.1 \\ 
LMC0543 & $5\uph30\upm37\zdot\ups44$ & $-69\arcd46\arcm42\zdot\arcs9$&   14  & 8.0 & 0.1 \\ 
LMC0544 & $5\uph30\upm39\zdot\ups41$ & $-69\arcd51\arcm12\zdot\arcs0$&   25  & 7.6 & 0.05 \\ 
LMC0545 & $5\uph30\upm39\zdot\ups55$ & $-70\arcd13\arcm06\zdot\arcs9$&   11  & 8.9 & 0.1 \\ 
\hline}

\setcounter{table}{0}
\MakeTableSep{|c|c|c|c|c|c|}{12.5cm}{continued}%10
{\hline
Name  &  $\alpha_{2000}$ & $\delta_{2000}$ & Radius
& $\log t$ & $\sigma_{\log t}$\\
OGLE-CL-& & & [\arcs] & & \\
\hline\xrule
LMC0546 & $5\uph30\upm40\zdot\ups70$ & $-70\arcd13\arcm21\zdot\arcs2$&   14  & 8.9 & 0.05 \\ 
LMC0547 & $5\uph30\upm42\zdot\ups81$ & $-69\arcd39\arcm01\zdot\arcs7$&   25  & 7.7 & 0.1 \\ 
LMC0549 & $5\uph30\upm50\zdot\ups88$ & $-69\arcd25\arcm39\zdot\arcs5$&   20  & 6.9 & 0.05 \\ 
LMC0551 & $5\uph30\upm58\zdot\ups66$ & $-69\arcd57\arcm20\zdot\arcs7$&   21  & 8.0 & 0.1 \\ 
LMC0552 & $5\uph31\upm04\zdot\ups31$ & $-70\arcd10\arcm00\zdot\arcs6$&   25  & 8.3 & 0.05 \\ 
LMC0554 & $5\uph31\upm16\zdot\ups34$ & $-69\arcd37\arcm57\zdot\arcs0$&    8  & 8.0 & 0.1 \\ 
LMC0556 & $5\uph31\upm19\zdot\ups97$ & $-70\arcd12\arcm54\zdot\arcs1$&   29  & 8.15 & 0.05 \\ 
LMC0557 & $5\uph31\upm28\zdot\ups74$ & $-70\arcd05\arcm15\zdot\arcs9$&   27  & 8.1 & 0.1 \\ 
LMC0558 & $5\uph31\upm30\zdot\ups77$ & $-70\arcd01\arcm24\zdot\arcs5$&   18  & 8.8 & 0.1 \\ 
LMC0559 & $5\uph31\upm35\zdot\ups05$ & $-69\arcd56\arcm43\zdot\arcs4$&   31  & 8.2 & 0.1 \\ 
LMC0560 & $5\uph31\upm36\zdot\ups00$ & $-69\arcd39\arcm18\zdot\arcs5$&   25  & 7.6 & 0.05 \\ 
LMC0563 & $5\uph31\upm45\zdot\ups81$ & $-70\arcd15\arcm09\zdot\arcs5$&   13  & 7.45 & 0.05 \\ 
LMC0564 & $5\uph31\upm50\zdot\ups36$ & $-70\arcd17\arcm21\zdot\arcs5$&   20  & 8.4 & 0.1 \\ 
LMC0565 & $5\uph31\upm56\zdot\ups48$ & $-70\arcd09\arcm32\zdot\arcs5$&   49  & $>$9.2 & -- \\ 
LMC0566 & $5\uph32\upm01\zdot\ups06$ & $-70\arcd10\arcm42\zdot\arcs6$&   21  & 8.2 & 0.1 \\ 
LMC0567 & $5\uph32\upm11\zdot\ups72$ & $-69\arcd29\arcm41\zdot\arcs1$&   25  & 8.2 & 0.1 \\ 
LMC0569 & $5\uph32\upm13\zdot\ups77$ & $-70\arcd02\arcm00\zdot\arcs3$&   23  & 8.0 & 0.1 \\ 
LMC0570 & $5\uph32\upm31\zdot\ups68$ & $-69\arcd34\arcm59\zdot\arcs7$&   24  & 8.5 & 0.08 \\ 
LMC0571 & $5\uph32\upm35\zdot\ups53$ & $-70\arcd00\arcm19\zdot\arcs8$&   24  & 8.5 & 0.1 \\ 
LMC0574 & $5\uph32\upm46\zdot\ups01$ & $-69\arcd52\arcm04\zdot\arcs6$&   14  & 8.3 & 0.1 \\ 
LMC0575 & $5\uph32\upm47\zdot\ups49$ & $-69\arcd39\arcm16\zdot\arcs2$&   26  & 9.0 & 0.1 \\ 
LMC0576 & $5\uph32\upm48\zdot\ups76$ & $-70\arcd26\arcm07\zdot\arcs4$&   10  & $>$8.9 & -- \\ 
LMC0577 & $5\uph32\upm48\zdot\ups86$ & $-70\arcd27\arcm23\zdot\arcs0$&   25  & 8.2 & 0.05 \\ 
LMC0578 & $5\uph32\upm51\zdot\ups25$ & $-70\arcd26\arcm01\zdot\arcs5$&   14  & 8.65 & 0.05 \\ 
LMC0579 & $5\uph32\upm57\zdot\ups04$ & $-69\arcd57\arcm06\zdot\arcs8$&   12  & 8.3 & 0.1 \\ 
LMC0580 & $5\uph32\upm58\zdot\ups90$ & $-70\arcd08\arcm24\zdot\arcs1$&   14  & 8.9 & 0.1 \\ 
LMC0582 & $5\uph33\upm04\zdot\ups68$ & $-70\arcd30\arcm46\zdot\arcs8$&   20  & 8.05 & 0.05 \\ 
LMC0583 & $5\uph33\upm06\zdot\ups18$ & $-70\arcd02\arcm30\zdot\arcs8$&   23  & 8.3 & 0.1 \\ 
LMC0585 & $5\uph33\upm21\zdot\ups91$ & $-69\arcd57\arcm20\zdot\arcs6$&   39  & 8.05 & 0.05 \\ 
LMC0586 & $5\uph33\upm23\zdot\ups04$ & $-70\arcd01\arcm48\zdot\arcs8$&   30  & 7.4 & 0.1 \\ 
LMC0589 & $5\uph33\upm26\zdot\ups40$ & $-70\arcd06\arcm51\zdot\arcs9$&   11  & 8.0 & 0.2 \\ 
LMC0590 & $5\uph33\upm39\zdot\ups71$ & $-70\arcd08\arcm40\zdot\arcs9$&   12  & 9.05 & 0.05 \\ 
LMC0591 & $5\uph33\upm40\zdot\ups98$ & $-69\arcd54\arcm58\zdot\arcs1$&   29  & 8.2 & 0.08 \\ 
LMC0592 & $5\uph33\upm41\zdot\ups10$ & $-69\arcd51\arcm43\zdot\arcs7$&   31  & 8.9 & 0.05 \\ 
LMC0594 & $5\uph33\upm48\zdot\ups46$ & $-69\arcd57\arcm03\zdot\arcs6$&   27  & 8.35 & 0.08 \\ 
LMC0595 & $5\uph33\upm49\zdot\ups50$ & $-69\arcd53\arcm18\zdot\arcs1$&   22  & 8.6 & 0.1 \\ 
LMC0596 & $5\uph33\upm57\zdot\ups53$ & $-69\arcd38\arcm38\zdot\arcs8$&   29  & 7.6 & 0.1 \\ 
LMC0597 & $5\uph33\upm57\zdot\ups85$ & $-70\arcd14\arcm32\zdot\arcs5$&   25  & 8.9 & 0.1 \\ 
LMC0598 & $5\uph34\upm00\zdot\ups48$ & $-69\arcd40\arcm21\zdot\arcs8$&   30  & 8.5 & 0.1 \\ 
LMC0600 & $5\uph34\upm07\zdot\ups80$ & $-69\arcd55\arcm20\zdot\arcs0$&   23  & 7.9 & 0.05 \\ 
LMC0602 & $5\uph34\upm30\zdot\ups86$ & $-69\arcd46\arcm50\zdot\arcs3$&   20  & 6.7 & 0.1 \\ 
LMC0603 & $5\uph34\upm31\zdot\ups91$ & $-70\arcd03\arcm57\zdot\arcs0$&   44  & 8.1 & 0.05 \\ 
LMC0604 & $5\uph34\upm38\zdot\ups36$ & $-69\arcd41\arcm36\zdot\arcs0$&   19  & 8.2 & 0.2 \\ 
LMC0605 & $5\uph34\upm40\zdot\ups36$ & $-69\arcd44\arcm50\zdot\arcs1$&   24  & 6.7 & 0.05 \\ 
LMC0606 & $5\uph34\upm40\zdot\ups93$ & $-70\arcd11\arcm41\zdot\arcs5$&    8  & 9.0 & 0.1 \\ 
LMC0607 & $5\uph34\upm42\zdot\ups06$ & $-70\arcd33\arcm42\zdot\arcs5$&   37  & 8.1 & 0.08 \\ 
LMC0608 & $5\uph34\upm46\zdot\ups65$ & $-69\arcd44\arcm35\zdot\arcs2$&   23  & 8.05 & 0.05 \\ 
LMC0609 & $5\uph34\upm50\zdot\ups49$ & $-69\arcd54\arcm58\zdot\arcs5$&   31  & 8.0 & 0.1 \\ 
LMC0610 & $5\uph34\upm56\zdot\ups00$ & $-69\arcd43\arcm07\zdot\arcs8$&   10  & 6.7 & 0.1 \\ 
LMC0611 & $5\uph34\upm59\zdot\ups58$ & $-70\arcd02\arcm34\zdot\arcs0$&   16  & 8.7 & 0.1 \\ 
\hline}

\setcounter{table}{0}
\MakeTableSep{|c|c|c|c|c|c|}{12.5cm}{continued}%11
{\hline
Name  &  $\alpha_{2000}$ & $\delta_{2000}$ & Radius
& $\log t$ & $\sigma_{\log t}$\\
OGLE-CL-& & & [\arcs] & & \\
\hline\xrule
LMC0612 & $5\uph35\upm02\zdot\ups55$ & $-70\arcd06\arcm35\zdot\arcs4$&   16  & 8.7 & 0.1 \\ 
LMC0613 & $5\uph35\upm03\zdot\ups46$ & $-70\arcd09\arcm30\zdot\arcs2$&   33  & 8.2 & 0.1 \\ 
LMC0614 & $5\uph35\upm04\zdot\ups06$ & $-70\arcd21\arcm30\zdot\arcs0$&   16  & 8.6 & 0.1 \\ 
LMC0615 & $5\uph35\upm07\zdot\ups71$ & $-70\arcd19\arcm32\zdot\arcs2$&   29  & 8.75 & 0.05 \\ 
LMC0616 & $5\uph35\upm14\zdot\ups02$ & $-69\arcd54\arcm21\zdot\arcs2$&   24  & 6.9 & 0.2 \\ 
LMC0618 & $5\uph35\upm23\zdot\ups52$ & $-69\arcd44\arcm41\zdot\arcs5$&   20  & 7.0,8.5 & 0.1,0.05 \\ 
LMC0619 & $5\uph35\upm30\zdot\ups68$ & $-70\arcd20\arcm56\zdot\arcs9$&   14  & 8.5 & 0.1 \\ 
LMC0621 & $5\uph35\upm36\zdot\ups65$ & $-70\arcd22\arcm11\zdot\arcs2$&   14  & 9.0 & 0.1 \\ 
LMC0622 & $5\uph35\upm38\zdot\ups66$ & $-70\arcd14\arcm23\zdot\arcs1$&   30  & 8.05 & 0.05 \\ 
LMC0623 & $5\uph35\upm49\zdot\ups30$ & $-69\arcd51\arcm10\zdot\arcs5$&   16  & 9.0 & 0.1 \\ 
LMC0624 & $5\uph35\upm50\zdot\ups81$ & $-69\arcd52\arcm35\zdot\arcs0$&   26  & 7.8 & 0.05 \\ 
LMC0626 & $5\uph35\upm54\zdot\ups42$ & $-70\arcd11\arcm28\zdot\arcs9$&   15  & 8.05 & 0.05 \\ 
LMC0627 & $5\uph35\upm56\zdot\ups66$ & $-70\arcd04\arcm23\zdot\arcs1$&   23  & 8.9 & 0.05 \\ 
LMC0628 & $5\uph35\upm58\zdot\ups46$ & $-70\arcd09\arcm17\zdot\arcs0$&   30  & 8.9 & 0.08 \\ 
LMC0629 & $5\uph36\upm11\zdot\ups69$ & $-70\arcd14\arcm27\zdot\arcs4$&   10  & 7.0 & 0.2 \\ 
LMC0630 & $5\uph36\upm22\zdot\ups58$ & $-70\arcd07\arcm40\zdot\arcs7$&   20  & 8.1 & 0.01 \\ 
LMC0631 & $5\uph36\upm33\zdot\ups95$ & $-70\arcd09\arcm55\zdot\arcs0$&   29  & 7.8 & 0.1 \\ 
LMC0632 & $5\uph36\upm53\zdot\ups68$ & $-70\arcd06\arcm21\zdot\arcs3$&   26  & 8.05 & 0.05 \\ 
LMC0633 & $5\uph36\upm54\zdot\ups52$ & $-70\arcd09\arcm43\zdot\arcs7$&   53  & 7.8 & 0.1 \\ 
LMC0634 & $5\uph36\upm56\zdot\ups21$ & $-70\arcd16\arcm10\zdot\arcs3$&   19  & 8.1 & 0.1 \\ 
LMC0635 & $5\uph36\upm56\zdot\ups83$ & $-69\arcd55\arcm19\zdot\arcs8$&   14  & 9.0 & 0.1 \\ 
LMC0636 & $5\uph37\upm01\zdot\ups53$ & $-70\arcd07\arcm36\zdot\arcs7$&   24  & 8.15 & 0.05 \\ 
LMC0637 & $5\uph37\upm13\zdot\ups54$ & $-70\arcd01\arcm25\zdot\arcs4$&   14  & $>$9.0 & -- \\ 
LMC0638 & $5\uph37\upm15\zdot\ups39$ & $-69\arcd53\arcm44\zdot\arcs7$&   18  & 8.65 & 0.05 \\ 
LMC0639 & $5\uph37\upm18\zdot\ups90$ & $-70\arcd09\arcm19\zdot\arcs0$&   16  & 7.8 & 0.1 \\ 
LMC0640 & $5\uph37\upm21\zdot\ups73$ & $-69\arcd53\arcm40\zdot\arcs5$&   13  & 8.9 & 0.1 \\ 
LMC0641 & $5\uph37\upm22\zdot\ups08$ & $-69\arcd58\arcm21\zdot\arcs2$&   34  & 8.85 & 0.05 \\ 
LMC0644 & $5\uph37\upm25\zdot\ups84$ & $-70\arcd13\arcm28\zdot\arcs6$&   16  & 8.4,8.7 & 0.1 \\ 
LMC0648 & $5\uph37\upm37\zdot\ups81$ & $-70\arcd13\arcm56\zdot\arcs4$&   59  & 8.1 & 0.05 \\ 
LMC0650 & $5\uph37\upm39\zdot\ups15$ & $-70\arcd08\arcm43\zdot\arcs9$&   37  & $>$8.3 & -- \\ 
LMC0651 & $5\uph37\upm42\zdot\ups36$ & $-70\arcd09\arcm54\zdot\arcs0$&   27  & 8.1 & 0.05 \\ 
LMC0654 & $5\uph38\upm10\zdot\ups92$ & $-69\arcd49\arcm50\zdot\arcs8$&   14  & 9.0 & 0.05 \\ 
LMC0655 & $5\uph38\upm21\zdot\ups26$ & $-70\arcd41\arcm06\zdot\arcs1$&   25  & 7.1 & 0.1 \\ 
LMC0656 & $5\uph38\upm24\zdot\ups10$ & $-70\arcd14\arcm00\zdot\arcs7$&   18  & 8.2 & 0.1 \\ 
LMC0657 & $5\uph38\upm26\zdot\ups88$ & $-70\arcd36\arcm30\zdot\arcs1$&   14  & 8.5 & 0.1 \\ 
LMC0659 & $5\uph38\upm33\zdot\ups26$ & $-69\arcd59\arcm30\zdot\arcs1$&    9  & 8.2 & 0.1 \\ 
LMC0661 & $5\uph38\upm49\zdot\ups73$ & $-70\arcd28\arcm30\zdot\arcs9$&   20  & 8.2 & 0.1 \\ 
LMC0662 & $5\uph38\upm53\zdot\ups40$ & $-69\arcd51\arcm46\zdot\arcs6$&   15  & 9.0 & 0.1 \\ 
LMC0664 & $5\uph39\upm00\zdot\ups27$ & $-69\arcd59\arcm19\zdot\arcs5$&   22  & 8.35 & 0.08 \\ 
LMC0665 & $5\uph39\upm05\zdot\ups63$ & $-70\arcd13\arcm46\zdot\arcs9$&   18  & 8.85 & 0.1 \\ 
LMC0668 & $5\uph39\upm28\zdot\ups93$ & $-70\arcd29\arcm44\zdot\arcs4$&   16  & 8.4 & 0.1 \\ 
LMC0670 & $5\uph39\upm36\zdot\ups01$ & $-69\arcd54\arcm28\zdot\arcs2$&   11  & 8.8 & 0.1 \\ 
LMC0671 & $5\uph39\upm36\zdot\ups49$ & $-70\arcd40\arcm27\zdot\arcs1$&   22  & 8.85 & 0.1 \\ 
LMC0672 & $5\uph39\upm39\zdot\ups32$ & $-70\arcd42\arcm45\zdot\arcs8$&   22  & 9.0 & 0.08 \\ 
LMC0673 & $5\uph39\upm45\zdot\ups67$ & $-70\arcd17\arcm01\zdot\arcs0$&   19  & 8.35 & 0.05 \\ 
LMC0674 & $5\uph39\upm50\zdot\ups25$ & $-70\arcd30\arcm51\zdot\arcs2$&   15  & 8.2 & 0.1 \\ 
LMC0675 & $5\uph39\upm58\zdot\ups20$ & $-70\arcd27\arcm17\zdot\arcs6$&   20  & 8.2 & 0.1 \\ 
LMC0676 & $5\uph40\upm14\zdot\ups14$ & $-70\arcd51\arcm25\zdot\arcs4$&   16  & 6.9 & 0.1 \\ 
LMC0678 & $5\uph40\upm39\zdot\ups91$ & $-70\arcd38\arcm26\zdot\arcs9$&   11  & 8.3 & 0.1 \\ 
LMC0679 & $5\uph40\upm56\zdot\ups60$ & $-70\arcd51\arcm27\zdot\arcs7$&   18  & 9.0 & 0.1 \\ 
\hline}

\setcounter{table}{0}
\MakeTableSep{|c|c|c|c|c|c|}{12.5cm}{continued}%12
{\hline
Name  &  $\alpha_{2000}$ & $\delta_{2000}$ & Radius
& $\log t$ & $\sigma_{\log t}$\\
OGLE-CL-& & & [\arcs] & & \\
\hline\xrule
LMC0680 & $5\uph41\upm01\zdot\ups85$ & $-70\arcd50\arcm50\zdot\arcs2$&   14  & 8.3 & 0.1 \\ 
LMC0681 & $5\uph41\upm04\zdot\ups85$ & $-70\arcd23\arcm19\zdot\arcs9$&   29  & 7.0 & 0.1 \\ 
LMC0682 & $5\uph41\upm07\zdot\ups31$ & $-70\arcd13\arcm03\zdot\arcs7$&   20  & 8.6 & 0.05 \\ 
LMC0683 & $5\uph41\upm13\zdot\ups41$ & $-70\arcd16\arcm35\zdot\arcs9$&   20  & 8.55 & 0.05 \\ 
LMC0685 & $5\uph41\upm21\zdot\ups79$ & $-70\arcd11\arcm44\zdot\arcs1$&   16  & 8.1 & 0.1 \\ 
LMC0686 & $5\uph41\upm29\zdot\ups28$ & $-70\arcd13\arcm58\zdot\arcs0$&   14  & 8.0 & 0.1 \\ 
LMC0690 & $5\uph41\upm57\zdot\ups00$ & $-70\arcd50\arcm54\zdot\arcs9$&   25  & 8.3 & 0.1 \\ 
LMC0691 & $5\uph42\upm17\zdot\ups59$ & $-70\arcd39\arcm43\zdot\arcs1$&   13  & 8.7 & 0.05 \\ 
LMC0692 & $5\uph42\upm32\zdot\ups42$ & $-70\arcd38\arcm04\zdot\arcs7$&    9  & $>$8.7 & -- \\ 
LMC0693 & $5\uph42\upm37\zdot\ups45$ & $-69\arcd57\arcm05\zdot\arcs7$&   18  & 8.85 & 0.08 \\ 
LMC0694 & $5\uph42\upm40\zdot\ups74$ & $-70\arcd29\arcm43\zdot\arcs6$&   28  & 8.5 & 0.1 \\ 
LMC0695 & $5\uph42\upm46\zdot\ups87$ & $-70\arcd09\arcm43\zdot\arcs2$&   18  & 9.05 & 0.05 \\ 
LMC0696 & $5\uph42\upm55\zdot\ups34$ & $-70\arcd39\arcm19\zdot\arcs4$&   13  & 8.65 & 0.05 \\ 
LMC0697 & $5\uph43\upm06\zdot\ups63$ & $-70\arcd16\arcm37\zdot\arcs9$&   23  & 8.7 & 0.1 \\ 
LMC0698 & $5\uph43\upm08\zdot\ups35$ & $-70\arcd24\arcm59\zdot\arcs4$&   16  & 8.9 & 0.05 \\ 
LMC0699 & $5\uph43\upm09\zdot\ups96$ & $-70\arcd34\arcm16\zdot\arcs4$&   23  & 8.7 & 0.05 \\ 
LMC0700 & $5\uph43\upm12\zdot\ups72$ & $-70\arcd38\arcm23\zdot\arcs3$&   42  & 8.4 & 0.05 \\ 
LMC0701 & $5\uph43\upm18\zdot\ups20$ & $-70\arcd44\arcm35\zdot\arcs3$&   11  & 8.0 & 0.1 \\ 
LMC0702 & $5\uph43\upm34\zdot\ups29$ & $-70\arcd30\arcm39\zdot\arcs3$&   16  & 8.85 & 0.05 \\ 
LMC0703 & $5\uph43\upm38\zdot\ups33$ & $-70\arcd33\arcm56\zdot\arcs1$&   18  & 8.0 & 0.1 \\ 
LMC0704 & $5\uph43\upm41\zdot\ups52$ & $-70\arcd36\arcm30\zdot\arcs3$&   23  & 9.0 & 0.05 \\ 
LMC0705 & $5\uph43\upm45\zdot\ups47$ & $-70\arcd51\arcm15\zdot\arcs1$&   11  & 9.0 & 0.1 \\ 
LMC0706 & $5\uph43\upm45\zdot\ups95$ & $-70\arcd53\arcm28\zdot\arcs4$&   19  & 8.9 & 0.1 \\ 
LMC0707 & $5\uph43\upm47\zdot\ups18$ & $-70\arcd07\arcm55\zdot\arcs1$&   12  & 8.7 & 0.1 \\ 
LMC0708 & $5\uph43\upm55\zdot\ups64$ & $-70\arcd36\arcm37\zdot\arcs6$&   11  & 9.0 & 0.1 \\ 
LMC0709 & $5\uph43\upm56\zdot\ups57$ & $-70\arcd55\arcm40\zdot\arcs2$&   17  & 7.9 & 0.1 \\ 
LMC0710 & $5\uph44\upm06\zdot\ups00$ & $-70\arcd31\arcm58\zdot\arcs3$&   17  & 8.3 & 0.2 \\ 
LMC0711 & $5\uph44\upm14\zdot\ups10$ & $-70\arcd39\arcm19\zdot\arcs8$&   16  & 8.3 & 0.1 \\ 
LMC0712 & $5\uph44\upm14\zdot\ups50$ & $-70\arcd40\arcm09\zdot\arcs5$&   20  & 8.3 & 0.1 \\ 
LMC0714 & $5\uph44\upm22\zdot\ups18$ & $-70\arcd15\arcm22\zdot\arcs2$&   17  & 9.0 & 0.1 \\ 
LMC0715 & $5\uph44\upm33\zdot\ups07$ & $-70\arcd59\arcm35\zdot\arcs3$&   32  & 8.2 & 0.05 \\ 
LMC0718 & $5\uph44\upm44\zdot\ups66$ & $-71\arcd00\arcm21\zdot\arcs3$&    8  & 8.4 & 0.1 \\ 
LMC0719 & $5\uph44\upm46\zdot\ups18$ & $-70\arcd17\arcm16\zdot\arcs6$&   14  & 8.2 & 0.1 \\ 
LMC0720 & $5\uph44\upm47\zdot\ups26$ & $-70\arcd24\arcm21\zdot\arcs9$&   11  & 8.2 & 0.2 \\ 
LMC0722 & $5\uph44\upm58\zdot\ups23$ & $-70\arcd13\arcm03\zdot\arcs3$&   17  & 9.0 & 0.05 \\ 
LMC0723 & $5\uph45\upm01\zdot\ups34$ & $-70\arcd32\arcm34\zdot\arcs2$&   20  & 9.0 & 0.1 \\ 
LMC0725 & $5\uph45\upm05\zdot\ups01$ & $-70\arcd14\arcm29\zdot\arcs4$&   27  & 8.4 & 0.05 \\ 
LMC0726 & $5\uph45\upm11\zdot\ups83$ & $-70\arcd43\arcm26\zdot\arcs7$&   27  & 8.3 & 0.08 \\ 
LMC0728 & $5\uph45\upm25\zdot\ups13$ & $-70\arcd24\arcm03\zdot\arcs9$&   18  & 8.6 & 0.1 \\ 
LMC0729 & $5\uph45\upm31\zdot\ups62$ & $-70\arcd45\arcm33\zdot\arcs7$&   23  & 8.6 & 0.1 \\ 
LMC0731 & $5\uph45\upm46\zdot\ups36$ & $-70\arcd43\arcm09\zdot\arcs0$&   16  & 8.7 & 0.1 \\ 
LMC0732 & $5\uph45\upm59\zdot\ups18$ & $-70\arcd43\arcm45\zdot\arcs8$&    9  & 7.8 & 0.2 \\ 
LMC0736 & $5\uph46\upm41\zdot\ups10$ & $-70\arcd50\arcm51\zdot\arcs8$&   11  & 8.25 & 0.08 \\ 
LMC0737 & $5\uph46\upm47\zdot\ups18$ & $-70\arcd49\arcm58\zdot\arcs5$&   15  & 8.35 & 0.05 \\ 
LMC0738 & $5\uph46\upm47\zdot\ups44$ & $-70\arcd35\arcm21\zdot\arcs0$&   11  & 8.3 & 0.05 \\ 
LMC0739 & $5\uph46\upm51\zdot\ups26$ & $-70\arcd30\arcm39\zdot\arcs9$&   19  & 8.0 & 0.15 \\ 
LMC0741 & $5\uph47\upm17\zdot\ups17$ & $-70\arcd48\arcm59\zdot\arcs1$&   23  & 8.9 & 0.1 \\ 
LMC0742 & $5\uph47\upm23\zdot\ups15$ & $-70\arcd26\arcm37\zdot\arcs4$&   24  & 8.8 & 0.05 \\ 
LMC0744 & $5\uph47\upm33\zdot\ups59$ & $-70\arcd57\arcm55\zdot\arcs8$&   23  & 8.85 & 0.05 \\ 
LMC0745 & $5\uph47\upm41\zdot\ups48$ & $-71\arcd12\arcm23\zdot\arcs9$&   15  & 8.7 & 0.05 \\ 
\hline}
\end{document}